\documentclass[a4paper,fleqn, usenatbib]{mnras}

\newcommand{\dynLogEinMass}{11.00\pm 0.02}
\newcommand{\dynAgreePerc}{5}
\newcommand{\ReArc}{3.49}
\newcommand{\ReKpc}{2.15}
\newcommand{\bestBH}{9.36}
\newcommand{\bestQ}{0.4949}
\newcommand{\bestP}{0.8390}
\newcommand{\bestU}{0.9910}
\newcommand{\bestTH}{88.05}
\newcommand{\bestPH}{75.76}

\newcommand{\bestDM}{21.00}
\newcommand{\bestDF}{2.00}
\newcommand{\bestML}{0.66}
\newcommand{\onsBH}{1.089}
\newcommand{\onsQ}{0.0090}
\newcommand{\onsP}{0.0281}
\newcommand{\onsU}{0.0026}
\newcommand{\onsDM}{2.67}
\newcommand{\onsDF}{0.094}
\newcommand{\onsML}{0.042}
\newcommand{\dynEinMass}{9.89 \pm 0.47}
\newcommand{\logsMassFOV}{11.23}
\newcommand{\logtMassFOV}{11.41}
\newcommand{\logsMassRe}{11.01}
\newcommand{\logtMassRe}{11.07}
\newcommand{\fDMRe}{10.67}
\newcommand{\meanTriFOV}{0.39}
\newcommand{\fOrbCold}{0.07}
\newcommand{\fOrbWarm}{0.47}
\newcommand{\fOrbHot}{0.39}
\newcommand{\fOrbCR}{0.07}
\newcommand{\totalSlope}{2.262 \pm 0.006}
\newcommand{\faceOnCS}{0.671}
\newcommand{\edgeOnCS}{1.165}
\newcommand{\lensDynMassRatios}{0.96}

\usepackage{textcomp, tensor, graphicx, tabu, multirow, url}
\usepackage[inline]{enumitem}
\usepackage[stable]{footmisc}

\usepackage{amsmath}
\usepackage{amssymb}

\usepackage{listings} % input code snippets
\usepackage{array} % for `\newcolumntype`
\usepackage{xifthen}

\newcolumntype{'}{!{\vrule width 2pt}}

\newcolumntype{L}[1]{>{\raggedright\let\newline\\\arraybackslash\hspace{0pt}}m{#1}}
\newcolumntype{C}[1]{>{\centering\let\newline\\\arraybackslash\hspace{0pt}}m{#1}}
\newcolumntype{R}[1]{>{\raggedleft\let\newline\\\arraybackslash\hspace{0pt}}m{#1}}

\newcommand{\atd}{ATLAS\(^{\rm 3D}\)}

\DeclareMathOperator{\Msun}{\si{M_\odot}}
\DeclareMathOperator{\Lsun}{\si{L_\odot}}
\newcommand{\logM}[1][]{\ifthenelse{\isempty{#1}}{\log_{10}(M/\Msun)}{\log_{10}(M_{#1}/\Msun)}}
\newcommand{\ml}[1][]{\ifthenelse{\isempty{#1}}{M/L}{M_{#1}/L}}
\newcommand{\mlStar}{\ifmmode{\ml[{\star}]}\else{\(\ml[{\star}]\)}\fi}
\newcommand{\Rein}{\ifmmode{\theta_{\rm Ein.}}\else{\(\theta_{\rm Ein.}\)}\fi}
\newcommand{\tfo}[1]{\texttt{#1}}
\newcommand{\tso}[1]{\textsc{#1}}
\newcommand{\shw}{Schwarzschild}
\newcommand{\SZ}{\ifmmode{{\rm S}0}\else{\({\rm S}0\)}\fi}
\newcommand{\mgeS}{MGE\(_\mu\)}
\newcommand{\mgeT}{MGE\(_\Sigma\)}

\newcommand{\ND}[1]{\(#1{\rm D}\)}
\newcommand{\lcdm}{\ifmmode{\Lambda{\rm CDM}}\else{\(\Lambda{\rm CDM}\)}\fi}
\newcommand{\eff}{\mathrm{e}}
\newcommand{\mfive}{\ifmmode{m_{0.5}}\else{\(m_{0.5}\)}\fi}
\newcommand{\mone}{\ifmmode{m_{1.0}}\else{\(m_{1.0}\)}\fi}

\newcommand{\chemZH}    {\nuclide{[Z/H]}}

\usepackage{tensor}
\let\scshape\relax % to avoid a warning
\DeclareRobustCommand\scshape{%
  \not@math@alphabet\scshape\relax
  \ifnum\pdf@strcmp{\f@family}{\familydefault}=\z@
    \fontfamily{lmr}%
  \fi
  \fontshape\scdefault\selectfont}
\makeatother

\LetLtxMacro{\oldsqrt}{\sqrt} % makes all sqrts closed
\renewcommand{\sqrt}[1][\ ]{%
  \def\DHLindex{#1}\mathpalette\DHLhksqrt}
\def\DHLhksqrt#1#2{%
  \setbox0=\hbox{$#1\oldsqrt[\DHLindex]{#2\,}$}\dimen0=\ht0
  \advance\dimen0-0.2\ht0
  \setbox2=\hbox{\vrule height\ht0 depth -\dimen0}%
  {\box0\lower0.71pt\box2}}

\usepackage{siunitx}
\DeclareSIUnit\parsec{pc}
\DeclareSIUnit\year{yr}
\DeclareSIUnit\angstrom{\text{Å}}
\usepackage{cleveref}
\crefname{section}{Section}{Sections}
\Crefname{section}{Section}{Sections}
\crefname{subsection}{Section}{Sections}
\Crefname{subsection}{Section}{Sections}
\crefname{subsubsection}{Section}{Sections}
\Crefname{subsubsection}{Section}{Sections}
\crefname{figure}{Fig.}{Figures}
\Crefname{figure}{Fig.}{Figures}
\crefname{table}{Table}{Tables}
\Crefname{table}{Table}{Tables}
\crefname{equation}{Eq.}{Equations}
\Crefname{equation}{Eq.}{Equations}

\graphicspath{ {figures//} }

\title[Orbital vs. Lensing Models]{Comparing Lensing and Stellar Orbital Models of a Nearby Massive Strong-Lens Galaxy}

\author[Poci \& Smith]{
Adriano Poci$^{1}$\thanks{E-mail: adriano.poci@durham.ac.uk},
Russell J. Smith$^{1}$
\\
$^{1}$Centre for Extragalactic Astronomy, University of Durham, Stockton Road, Durham DH1 3LE, United Kingdom}

\date{Accepted XXX. Received YYY; in original form ZZZ}

\pubyear{2021}

\begin{document}
\label{firstpage}
\pagerange{\pageref{firstpage}--\pageref{lastpage}}
\maketitle

% Abstract of the paper
\begin{abstract}
Exploiting the relative proximity of the nearby strong-lens galaxy SNL-1, we present a critical comparison of the mass estimates derived from independent modelling techniques. We fit triaxial orbit-superposition dynamical models to spatially-resolved stellar kinematics, and compare to the constraints derived from lens modelling of high-resolution photometry. From the dynamical model, we measure the total (dynamical) mass enclosed within a projected aperture of radius the Einstein radius to be \(\logM[{\rm Ein.}] = \dynLogEinMass\), which agrees with previous measurements from lens modelling to within \(\dynAgreePerc\%\). We then explore the intrinsic (de-projected) properties of the best-fitting dynamical model. We find that SNL-1 has approximately-constant, intermediate triaxiality at all radii. It is oblate-like in the inner regions (around the Einstein radius) and tends towards spherical at larger radii. The stellar velocity ellipsoid gradually transforms from isotropic in the very central regions to radially-biased in the outskirts. We find that SNL-1 is dynamically consistent with the broader galaxy population, as measured by the relative fraction of orbit `temperatures' compared to the CALIFA survey. On the mass--size plane, SNL-1 occupies the most-compact edge given its mass, compared to both the MaNGA and SAMI surveys. Finally, we explore how the observed lensing configuration is affected by the orientation of the lens galaxy. We discuss the implications of such detailed models on future combined lensing and dynamical analyses.
\end{abstract}

% Select between one and six entries from the list of approved keywords.
% Don't make up new ones.

%%%%%%%%%%%%%%%%%%%%%%%%%%%%%%%%%%%%%%%%%%%%%%%%%%
% ------------------------------------------------------------------------------

\section{Introduction}
\label{sec:intro}
The mass of a galaxy is one of the most critical properties controlling its evolution over cosmic time. This is evidenced by the plethora of observed correlations of other galactic properties with mass \citep[e.g.][]{gallazzi2005, cortese2014, barone2021, tian2021a}. Measuring the mass, however, is notoriously difficult for a number of reasons. Transforming the observed luminosity of a galaxy into a baryonic mass is prone to systematic uncertainties relating to distances, the underlying stellar populations, and complications such as dust. Moreover, a significant portion of a galaxy's mass, the dark matter (DM), is completely invisible and therefore must be inferred from its effect on luminous matter.\par%, through kinematics \citep[e.g.][]{cappellari2013, zhu2016, tortora2019} or gravitational lensing \citep[e.g.][]{auger2010}.\par
Direct probes of the gravitational potential can circumvent many of these issues, and can constrain the total (gravitational) mass, irrespective of the specific combination of baryonic and dark matter. To this end, gravitational lensing offers a relatively assumption-independent avenue for constraining the total enclosed mass within the characteristic Einstein radius \(\Rein\) \citep[e.g.][]{chae2003, treu2010a}. Moreover, in general, the confidence with which a lensing model can be constrained is dependent on the geometry and multiplicity of the images produced \citep[e.g.][]{shu2015, smith2018}. Finally, the main degeneracy in lensing models --- in particular those with only two lensed images and/or little spatial structure in the images --- is between the mass which is responsible for the lensing, and contributions from external effects (e.g. the degeneracy between mass and shear). However with high-quality data, this degeneracy can in principle be overcome by using the flux information in the pixels of the source images, rather than just their positions around the lens \citep{collier2018a}. Unfortunately lensing analyses are naturally limited to the small samples of galaxies which are acting as strong lenses to background sources \citep[e.g.][]{bolton2006}, and so mass census of the galaxy population using this technique is similarly limited.\par
Kinematics of the baryonic components can also provide robust estimates of the total enclosed mass of galaxies, since they directly trace the total gravitational potential (irrespective of the specific combination of baryons and DM). However, measuring kinematics requires spectroscopy with relatively high signal-to-noise \((S/N)\), which is considerably more expensive than photometry, especially at the redshift of typical strong-lens systems. This requirement tightens with increasing generality of the dynamical model employed.\par
Comparing the measurements of the projected enclosed mass for the same galaxy provides a critical test of these two techniques, and consequently whether previous works using either technique are directly comparable. Dynamical models have in fact been applied to a sample of lensed galaxies, finding that models with DM better match the lensing analyses compared to mass-follows-light models \citep{thomas2011}. Using the lens sample from the Sloan Lens ACS Survey \citep[SLACS;][]{bolton2006}, two-integral dynamical models were explored using spatially-resolved stellar kinematics \citep{czoske2008, barnabe2009, barnabe2011}. However, given the relatively large distances to the majority of strong lens galaxies (for instance, SLACS sample is at \(0.08 < z < 0.35\), with the most massive around \(z\sim 0.25\)), the available data make detailed dynamical modelling challenging. These analyses \citep[e.g.][]{treu2004, koopmans2009, auger2010, thomas2011} more often rely on mass estimators which require only a central velocity dispersion \citep[such as in][and their respective assumptions]{cappellari2006, wolf2010, campbell2017}. The sample of lenses collected as part of the SINFONI Nearby Elliptical Lens Locator Survey \citep[SNELLS;][]{smith2015a} is unique for the relatively small distances to the lens galaxies. \cite{newman2017} have exploited their proximity to derive combined constraints from stellar-population and lensing analyses for a subset of the SNELLS galaxies in order to measure potential variations of the stellar Initial Mass Function (IMF).\par
In this work, we also take advantage of the distances of the SNELLS galaxies to conduct a detailed dynamical analysis and comparison to constraints from lens modelling for SNL-1 (ESO286-G022). We compute triaxial three-integral orbit-based dynamical models of SNL-1 using the measured spatially-resolved kinematic moments which do not require specific assumptions regarding the mass distribution or orbital anisotropy. For consistency with the lens modelling of \cite{collier2018a} --- to which we directly compare the dynamical modelling results --- we assume the cosmology of the {\em Wilkinson Microwave Anisotropy Probe} ({\em WMAP}) \(7\)-year experiment \citep{komatsu2011}. Physical properties of SNL-1 are summarised in \cref{tab:props}.
\begin{table*}
\newcolumntype{L}{>{$}l<{$}} % math-mode version of "l" column type
\begin{tabular}{rL|L|c}\hline\hline
    Redshift & z & 0.0312 & \text{\cite{smith2015a}}\\\hline
    \multirow{2}{*}{Einstein radius} & \multirow{2}{*}{\(\Rein\)} & 2.38\si{\arcsecond} & \multirow{2}{*}{\cite{smith2015a}}\\
     && 1.48\ \si{\kilo\parsec} &\\\hline
    Total mass & \logM & 10.98 & \text{\cite{collier2018a}}\\\hline
    \multirow{2}{*}{Distance} & \multirow{2}{*}{\(D\)} & 132\ \si{\mega\parsec} & co-moving\\
     && 128\ \si{\mega\parsec} & angular diameter\\\hline
    \multirow{2}{*}{Effective radius} & \multirow{2}{*}{\(R_\eff^{F814W}\)} & \ReArc\si{\arcsecond} & \multirow{2}{*}{---}\\
     && \ReKpc\ \si{\kilo\parsec} & \\\hline
\end{tabular}
\caption{Physical properties of SNL-1 from the literature and this work.}
\label{tab:props}
\end{table*}
%This results in comoving and angular diameter distances of \(132.00\ \si{\mega\parsec}\) and \(128.02\ \si{\mega\parsec}\), respectively. The strong lensing of a background source at \(z=0.926\) produces an Einstein radius of \(\Rein=2.38\si{\arcsecond}\) \citep[\(1.48\ \si{\kilo\parsec}\);][]{smith2015a}. We measure an \(F814W-\)band effective radius of \(R_\eff = 3.41\si{\arcsecond}\ (2.12\ \si{\kilo\parsec})\) along the major axis and an integrated velocity dispersion within this radius of \(\sigma_\eff =  306.84\ \si{\kilo\metre\per\second}\).

\section{Data \& Target}\label{sec:data}
The dynamical model used in this work (\cref{sec:methods}) requires high-quality photometry and spectroscopy in order to be robustly constrained. We utilise Hubble Space Telescope ({\em HST}) Advanced Camera for Surveys ({\rm ACS}) photometry in order to derive a model for the projected surface brightness of SNL-1, which is used as the tracer distribution for the dynamical model. Very Large Telescope (VLT) Multi-Unit Spectroscopic Explorer (MUSE) integral-field unit (IFU) spectroscopy is used to measure the stellar kinematics and star-formation history. The stellar kinematics will constrain the dynamical model directly, while the stellar populations will be used to derive a stellar mass model from the surface brightness.
\subsection{Photometry}
\begin{figure}
    \centerline{
        \includegraphics[width=\columnwidth]{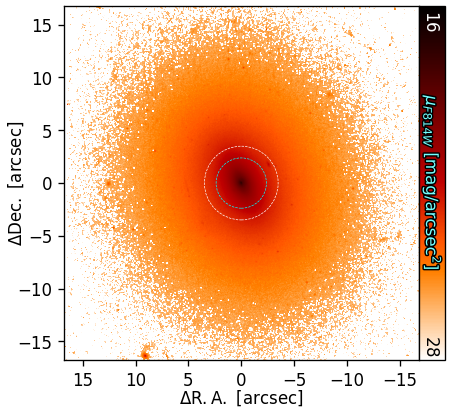}}
    \caption{\(F814W\) image of SNL-1. Circular apertures with radii \(R_\eff\) and \(\Rein\) are shown in white and cyan, respectively.}
    \label{img:photo}
\end{figure}

\cite{collier2018a} presented observations of SNL-1 in the \(F336W\) and \(F814W\) bands of {\em HST}/{\rm ACS}. From these data, we utilised the \(F814W\) band with an exposure time of \(1050\ \si{\second}\), as it is deeper than the \(F336W\) and consistent with the wavelength range of the spectroscopy. The \(F814W\) image is shown in \cref{img:photo}.
%We note that although our spectroscopy is truncated to \(6,700\si{\angstrom}\), the peak transmission of the \(F814W\) filter on ACS/WFC is at \(7,460\si{\angstrom}\)\footnote{according to the {\em HST} User Guide available at \href{https://etc.stsci.edu/etcstatic/users_guide/appendix_b_acs.html}{https://etc.stsci.edu/etcstatic/users\_guide/appendix\_b\_acs.html}}. In conjunction with the overall old ages present in SNL-1, we expect the difference in the underlying stellar populations being traced by the spectroscopy and photometry to be negligible.
We also have \(r\)-band data from FORS2 which we utilise for auxiliary investigations below. However, given the lower spatial resolution and non-uniform sky background of this data, we opt for the {\em HST} \(F814W\) for the main modelling of this work.
\subsection{Spectroscopy}\label{ssec:spec}
The spectroscopy of SNL-1 was obtained using MUSE in the wide-field mode (WFM) configuration without adaptive optics, with program ID 0100.B-0769(A). These data have an exposure time of \(1180\ \si{\second}\). They were reduced via the standard ESO MUSE pipeline. At the derived distance to SNL-1, the \(0.2\ \si{arcsec\per pixel}\) resolution of these data translate into a physical spatial resolution of \(124\ \si{\parsec}\), but with an estimated point-spread function (PSF) of \(\sim 1\si{\arcsecond}\ (\sim 621\ \si{\parsec})\) full-width half-maximum.\par
The MUSE data-cube was spatially binned to a target signal-to-noise \(S/N=50\) using a Python implementation\footnote{Available at \url{https://pypi.org/project/vorbin/}} of the Voronoi binning algorithm \citep{cappellari2003}. This \(S/N\) was chosen as a compromise between being able to robustly extract higher-order information from the spectral fits, and maintaining sufficient spatial sampling across the FOV in order to preserve the kinematic structures. For all data products, we consider the rest-frame spectral range \(\lambda \in [4700, 6700]\ \si{\angstrom}\), as red-ward of \(6700\ \si{\angstrom}\), the data are more noisy and contain sky emission lines. The binned data extend out to \(r_{\rm max}=9.67\si{\arcsecond}\ (6.01\ \si{\kilo\parsec})\) or \(\sim 2.7\ R_\eff\). To measure the stellar population properties, the binned spectra were fit using the {\rm E-MILES}\footref{fn:miles} single stellar population (SSP) templates \citep{vazdekis2016} within {\sc pPXF}\footnote{\label{fn:ppxf}Available at \url{https://pypi.org/project/ppxf/}} \citep{cappellari2008,cappellari2017}. Specifically, we used the models computed using the `BaSTI' isochrones \citep{pietrinferni2004}, with `base' (Solar neighbourhood) elemental abundances, and a fixed `revised' \cite{kroupa2001} IMF over the safe ranges of age \((t)\) and total metallicity \((\chemZH)\)\footnote{as defined at \url{http://research.iac.es/proyecto/miles/pages/ssp-models/safe-ranges.php}}. We included a multiplicative polynomial of order \(12\) to account for any mismatch between the absolute flux calibrations of the data and models, and any residual stellar continuum not accounted for by the models. A linear regularisation term \((\Delta=10)\) was also included, which favours a smooth SSP weight distribution in the underlying \(t-\chemZH\) space given otherwise degenerate solutions.\par
We used the resulting distribution of age and metallicity in each spatial bin to compute the \(F814W\)-band stellar mass-to-light ratio \((\ml[\star]_{F814W})\). For the stellar mass, the E-MILES predictions including the mass in stars and stellar remnants was used. To compute the luminosity, we used the SSP model spectra directly. First, we took each SSP template which was assigned non-zero weight in the fit, and computed the absolute magnitude by applying the \(F814W\) filter response curve redshifted to the distance of SNL-1. We then took the standard Solar spectrum of \cite{colina1996} from the {\em HST} CALSPEC database \citep{bohlin2014, bohlin2020} and computed its magnitude in the same way. In this way, we computed the total luminosity by weighting the individual SSP luminosities according to the weights of the spectral fit. The \ND{2} map of \(\ml[\star]_{F814W}\) is shown in the inset of \cref{img:ML}.\par
Measurements of the stellar kinematics were also made from the same data, in this instance fitting the spectra again in {\sc pPXF} using the {\rm MILES}\footnote{\label{fn:miles}Available at \url{http://research.iac.es/proyecto/miles/}} empirical stellar library \citep{sanchez-blazquez2006, falcon-barroso2011}. We employ stellar spectra for this fit because they have higher intrinsic resolution compared to the SSP models, which improves the fit to complex absorption-line shapes. In this instance, an additive polynomial of order \(12\) was included to ensure that the shapes of the absorption features are accurately reproduced, and no regularisation of the weights is imposed. In order to robustly constrain the dynamical model (\cref{sec:methods}), we measured the first four Gauss-Hermite moments of the line-of-sight velocity distribution (LOSVD), shown in the top row of \cref{img:schw}.\par
During the fit to every spectrum, bad spectral pixels were iteratively clipped. This efficiently masks emission (which is not accounted for in the fitting), any sky lines, and other spurious artefacts in the spectrum. This scheme would also account for any emission from the source galaxy, as well as template mismatch during the fitting. Since the lensed images are undetected in continuum, no spatial masking is required.

\section{Dynamical Model}\label{sec:methods}
In this work we aim to construct an accurate dynamical model in order to derive constraints which are in line with those produced from the lensing analysis. Since the latter is subject to few assumptions, we also wish to minimise those imposed on the dynamical model. We therefore apply a highly-general orbit-superposition technique built on the original premise of \cite{schwarzschild1979}. We utilise the triaxial three-integral implementation described in \cite{vandenbosch2008, vandeven2008, zhu2018a}, and validated on mock data in \cite{jin2019}. This technique allows for freedom in the shape of the velocity ellipsoid and intrinsic mass distribution. Confronting this method with the result from lensing provides a critical test of the ingredients that are required to fit the observational data. We first briefly describe those ingredients here.
\subsection{Projected Mass Model}\label{ssec:mge}
Dynamical models require a description of the mass distribution in order to compute kinematic predictions, but this is not a direct observable. In practise, the mass distribution is derived from some projected constraints, then deprojected into an intrinsic distribution via specific assumptions about the shape of the galaxy. In this work, we computed a model of the projected surface brightness by fitting a multi-Gaussian Expansion \citep[MGE;][]{monnet1992, emsellem1994} to the \(F814W\) photometry using a Python implementation\footnote{Available at \url{https://pypi.org/project/mgefit/}} \citep{cappellari2002}. To convert this fit into physical units we took into account the surface-brightness dimming due to redshift, Galactic extinction according to \cite{schlafly2011}, as well as a K-correction in the redshifted \(F814W\) band. This model describes the projected distribution of the visible tracer of the underlying gravitational potential. Since it describes the surface brightness, we refer to it is \mgeS.\par
\begin{figure}
    \centerline{
        \includegraphics[width=\columnwidth]{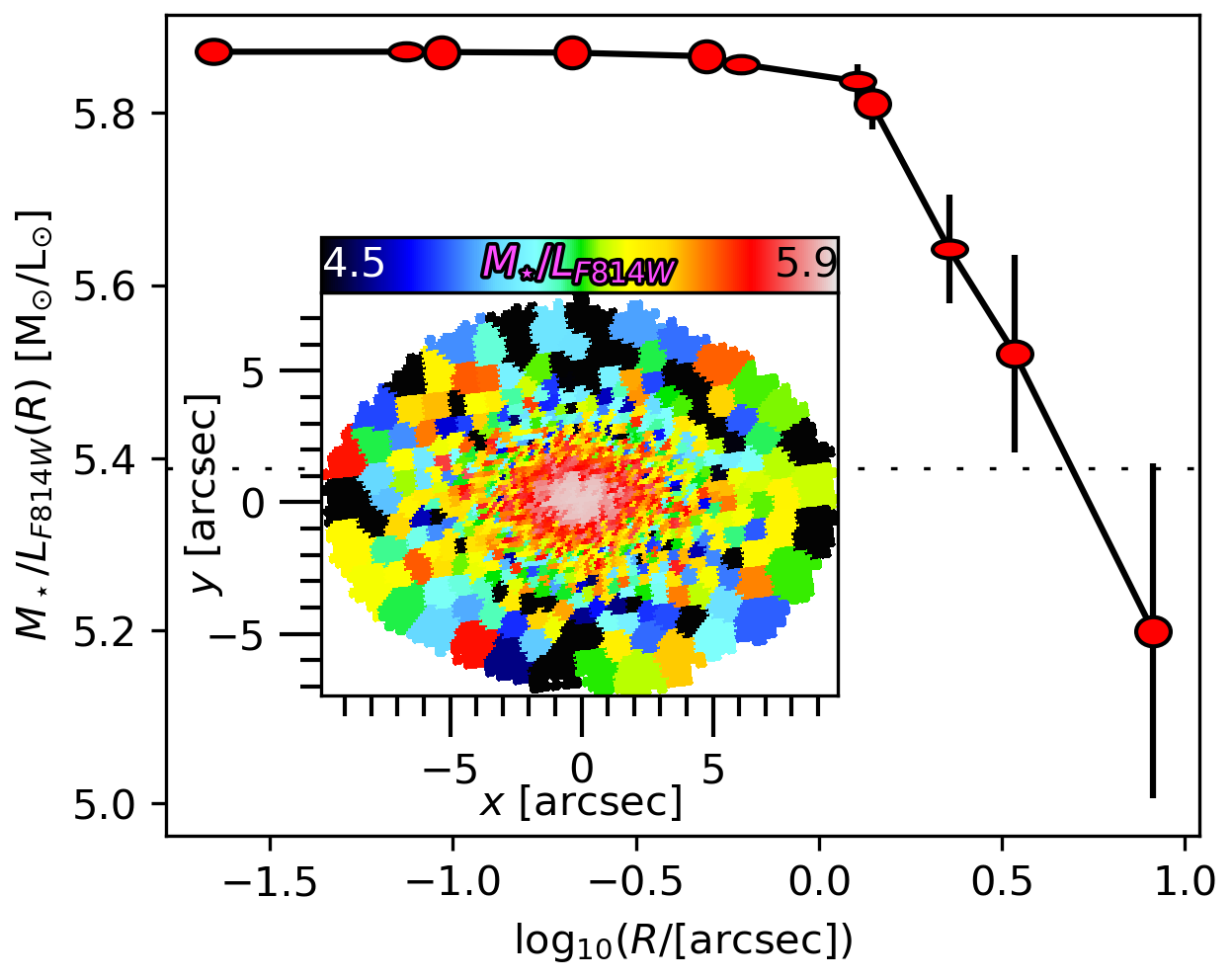}}
    \caption{{\em Main:} Azimuthally-sampled radial profile of the measured \(\ml[\star]_{F814W}\) from the full spectral fitting in \cref{ssec:spec}. The measurements are sampled in a thin annulus around the location (\(\sigma\)) of every Gaussian of \mgeS\ (see text). The data points depict the shape of the annulus taken for each Gaussian component, governed by its axis ratio. The vertical `error bars' illustrate the range of \mlStar\ within each annulus. The black dashed line shows the average value of \(\ml[\star]_{F814W}\) over the individual Voronoi bins. {\em Inset:} The spatially-resolved map of \(\ml[\star]_{F814W}\).}
    \label{img:ML}
\end{figure}

We still require, however, a description of the mass. To this end, we utilise the spectroscopic \(\ml[\star]_{F814W}\) in order to convert the stellar luminosity into mass. This is achieved following the approach of \cite{poci2017}, which we outline here \citep[with similar approaches from][]{li2017, mitzkus2017, yang2020}. Since \mgeS\ is constructed from the co-addition of overlapping \ND{2} Gaussians of different projected shapes, widths, and intensities, we derived the corresponding \mlStar\ for each Gaussian individually. For each Gaussian, we took a thin projected annulus centred at its `width' (radial extent) and with the same axis ratio, and calculated the mean \mlStar\ from the {\sc pPXF} stellar-population fit within that annulus. The mean value is then the factor which converts the luminosity of that Gaussian into mass. The final set of scaled Gaussians represents a projected mass model, which we denote \mgeT. In this implementation, the tailored azimuthal sampling naturally accounts for the interplay between the individual Gaussians and the changing shape of the light distribution with radius. This allows the dynamical model to account for the spatial variations of \mlStar\ driven by age and metallicity variations, notwithstanding the evidence for internal IMF variations in massive ETGs similar to SNL-1 \citep[e.g.][]{vandokkum2017, vaughan2018, labarbera2019}. Given the spheroidal geometry of SNL-1, we expect radial profiles to capture these variations. The \ND{2} map and resulting radial profile of the measured \(\ml[\star]_{F814W}\) are shown in \cref{img:ML}, and both \mgeS\ and \mgeT\ are tabulated in \cref{app:MGE}. With respect to the spatially-averaged value denoted in \cref{img:ML}, it can be seen that explicitly including the radial profile in our mass model accounts for the \(\sim \pm 10\%\) variations of \(\ml[\star]_{F814W}\) caused by age and metallicity gradients.\par
\subsection{Intrinsic Parameter Space Exploration}\label{ssec:schwParms}
The goal of the dynamical model is to find the best set of intrinsic (de-projected) properties which reproduce both the projected mass distribution and the observed (projected) stellar kinematics. The specific implementation of the \shw\ method used here describes the intrinsic mass distribution with seven parameters: \begin{enumerate}[label=(\emph{\alph*})] \item a parametrisation of the intrinsic shape of the stellar component with three axis ratios \(q=C/A\), \(p=B/A\), and \(u=A^\prime/A\), where \(A\), \(B\), and \(C\) are the intrinsic major, intermediate, and minor axes, respectively, and \(A^\prime\) is the projected major axis \item the mass of the central SMBH, \(M_\bullet\). This is implemented as a `dark' mass following a Plummer density profile \citep{dejonghe1987} \item two parameters describing the DM halo, assumed to be a spherical Navarro-Frenk-White (NFW) model \citep{navarro1996}. These are the concentration \(C_{\rm DM}\) and dark mass fraction at \(r_{200}\), \(f_{\rm DM}\) \item a global dynamical mass-to-light ratio. We denote this quantity \(\Upsilon\) to distinguish from the spectroscopic \mlStar. \(\Upsilon\) deepens or flattens the global gravitational potential, resulting in larger or smaller velocities, respectively, as needed to fit the kinematics. It can account for systematic effects caused by the absolute calibration of the SSP models, the particular choice of IMF, and/or the assumption of a spherical NFW DM halo.\end{enumerate}\par
For a single set of the above parameters, a large library of representative orbits are numerically integrated within the corresponding gravitational potential. The orbits conserve three integrals of motions: \(E\), \(I_2\), and \(I_3\). Our models sample these integrals in \(25\) logarithmic steps for \(E\), \(18\) linear steps for \(I_2\), and \(10\) linear steps for \(I_3\). To avoid discreteness (aliasing) in the models, each \((E, I_2, I_3)\) location was `dithered' \citep[as in][]{cappellari2006} by a factor of \(5\). This creates a cloud of orbits with adjacent initial conditions for every integration, alleviating any possible discreteness in the model observables without having to integrate five times as many orbits.\par
The complete orbit library was fit to the measured kinematics via a Non-Negative Least-Squares \citep[NNLS;][]{lawson1995} algorithm.
%In fact, we required the dynamical model to fit not only the measured kinematics up to the fourth Gauss-Hermite term, but also included maps of \(h5\) and \(h6\) which were zero everywhere but with non-zero error. In the absence of these higher-order maps, the implicit assumption is that the LOSVDs which are produces by the model in order to fit the measured kinematics are well-behaved beyond \(h4\), but with no indication as to whether or not this is actually the case. Including these higher terms explicitly permits mild variation of these moments \citep[c.f. the expected magnitude of variations in][]{liepold2020} but prevents large (and unknown) fluctuations.
As a boundary constraint, the orbits are also required to reproduce the projected luminosity (\mgeS). During this fit, a linear regularisation in imposed -- analogous to that used for spectral fitting -- in this case favouring smooth distributions in the orbital phase-space \((E, I_2, I_3)\). The result of the NNLS fit is a set of (luminosity) weights for all of the integrated orbits, whose weighted combination reproduces all of the kinematic observables.\par
\begin{figure*}
    \centerline{\includegraphics[width=\textwidth]{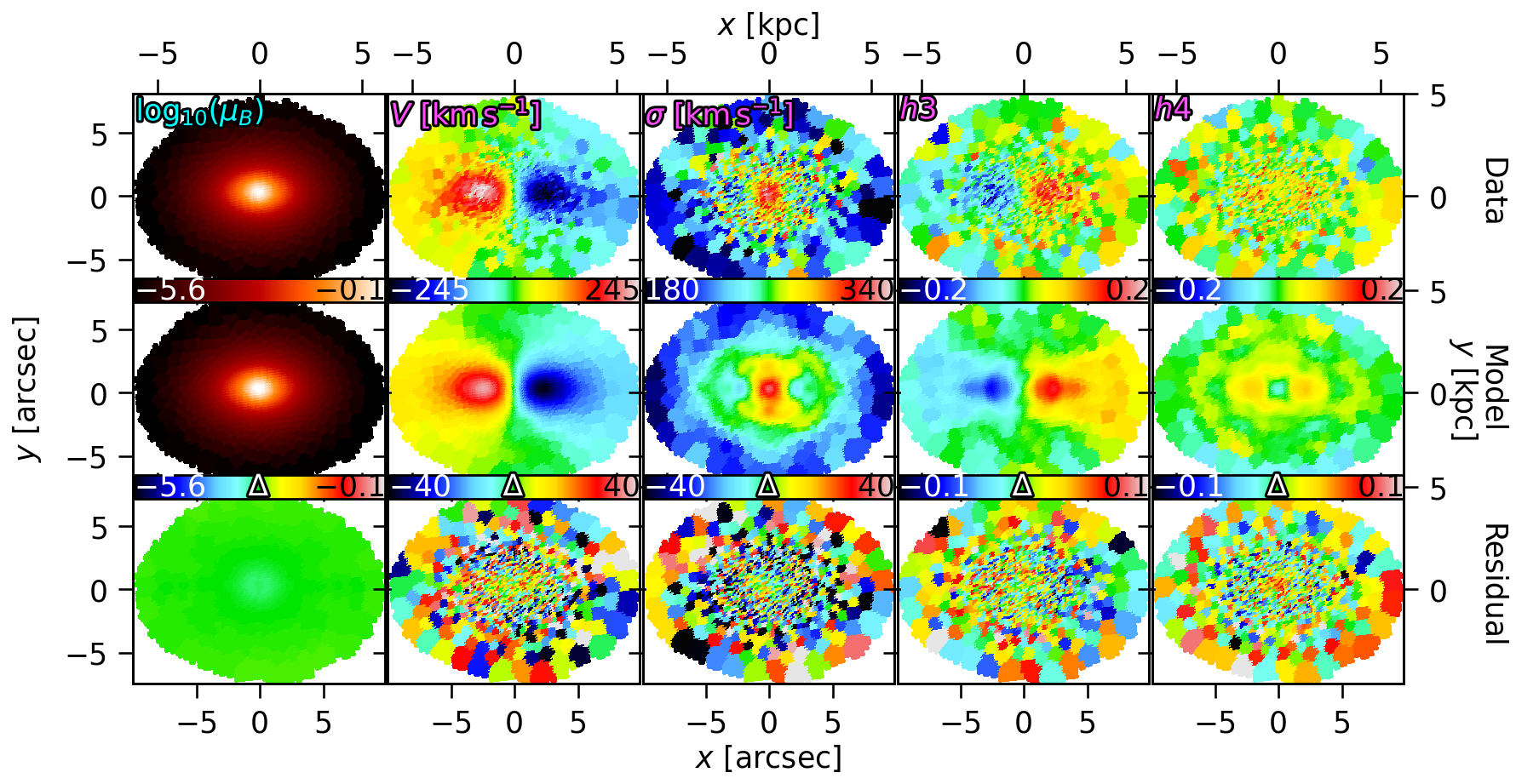}}
    \caption{\shw\ model fit to the measured kinematics of SNL-1. Data are shown in the top row and the model in the middle row. The residuals (\(\Delta = {\rm data} - {\rm model}\)) are shown in the bottom row, with their corresponding colour bars. From left to right, the columns contain the projected surface brightness, mean velocity, velocity dispersion, \(h3\), and \(h4\).}
    \label{img:schw}
\end{figure*}

It is then necessary to explore many possible sets of parameters -- different intrinsic gravitational potentials -- at each stage finding the best subset of orbits via NNLS.
%To efficiently conduct this computationally-expensive search, we fixed the mass of the SMBH to that derived from the \(M_\bullet-\sigma_\eff\) relation of \cite{kormendy2013}. For the measured \(\sigma_\eff\) this relation produces \(\logM[\bullet] = 9.20\). Using the relation of \cite{merritt2001b}, we estimate the radius of the sphere-of-influence for the given \(M_\bullet\) to be \(r = 0.12\si{\arcsecond}\) -- well below the MUSE PSF. We can therefore fix \(M_\bullet\) without impacting the accuracy of the model.
%Moreover, given the low fraction of DM expected for SNL-1, constraining both \(C_{\rm DM}\) and \(f_{\rm DM}\) with our relatively-central kinematic data would be challenging. We instead tied \(C_{\rm DM}\) to the value of \(f_{\rm DM}\) via the mass-concentration relation of \cite{dutton2014a}.
Each model consists of \(6\) gravitational parameters: \(M_\bullet\), \(q\), \(p\), \(u\), \(C_{\rm DM}\), and \(f_{\rm DM}\), where each set is evaluated in NNLS over a range of \(\Upsilon\) (since changing \(\Upsilon\) does not require re-integration of the orbits). We note here, however, that using the measured velocity dispersion within the effective radius, \(\sigma_\eff\), and the \(M_\bullet-\sigma_\eff\) relation of \cite{kormendy2013}, we estimate the radius of the sphere-of-influence for the given \(M_\bullet\) to be \(r = 0.12\si{\arcsecond}\) --- well below the MUSE PSF. Therefore, while we leave \(M_\bullet\) as a free parameter for completeness, we expect the constraints on it to be weak given our current data. The exploration of the parameter space is described in \cref{app:corner}. Overall, \(7233\) unique models were evaluated, and the best fit is shown in \cref{img:schw}.

\section{Results}
\begin{figure}
    \centerline{\includegraphics[width=\columnwidth]{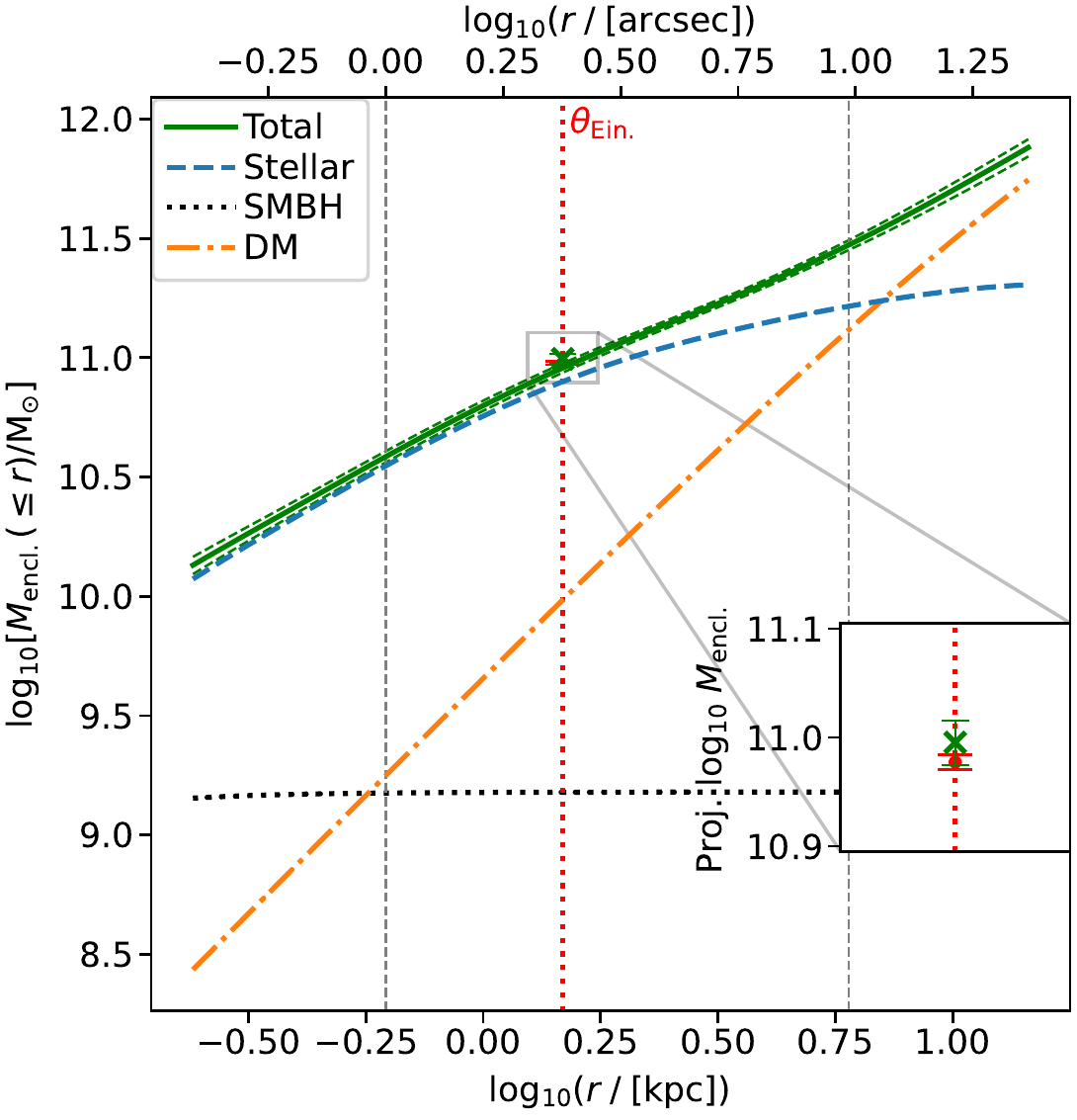}}
    \caption{Enclosed mass of SNL-1. Intrinsic de-projected profiles of the enclosed stellar and dark mass are shown in dashed blue and dot-dashed orange, respectively. The total enclosed mass is shown as the green solid line [given by \(M_\star(r) + M_{\rm DM}(r) + M_\bullet(r)\), although the SMBH is a point source over the radial range considered]. The black-hole mass is shown in dotted grey for reference. The green cross and the red errorbar illustrate the projected constraints from the dynamical and lensing models, respectively. The PSF and largest radial extent of the MUSE data are demarcated by the inner and outer vertical dashed lines, respectively. \(\Rein\) is shown as the red dotted line. For the dynamical model, the errorbar on the cross and the dashed green lines illustrate the spread of all models within \(1\sigma\) of the best-fitting solution. {\em Inset}: Zoom-in of the measurements of the projected Einstein mass from the lensing and dynamical models.}
    \label{img:profile}
\end{figure}

The parameters of the best-fitting \shw\ model are given in \cref{tab:FP}.
\begin{table}
\centerline{
\begin{tabu}{c|c|c|c}
    {\bf Parameter} & {\bf Description} & {\bf Best} & {\bf \(1\sigma\)}\\\hline\hline
    \(\log_{10}(M_\bullet/\Msun)\) & Black-Hole Mass & \(\bestBH\) & \(\onsBH\)\\\hline
    \(q\) & Intrinsic Shape & \(\bestQ\) & \(\onsQ\)\\
    \(p\) & Intrinsic Shape & \(\bestP\) & \(\onsP\)\\
    \(u\) & Intrinsic Shape & \(\bestU\) & \(\onsU\)\\\hline
    \(\theta^\prime\) & Viewing Angle & \(\bestTH\si{\degree}\)&\\
    \(\phi^\prime\) & Viewing Angle & \(\bestPH\si{\degree}\)&\\
    \(\psi^\prime\) & Viewing Angle & \(\bestPH\si{\degree}\)&\\\hline
    \(C_{\rm DM}\) & DM Concentration & \(\bestDM\) & \(\onsDM\)\\\hline
    \(\log_{10}\left[f_{\rm DM}\left(r_{200}\right)\right]\) & DM Fraction at \(r_{200}\) & \(\bestDF\) & \(\onsDF\)\\\hline
    \(\Upsilon\; [\si{\Msun/\Lsun}]\) & Global \(M/L\) & \(\bestML\) & \(\onsML\)
\end{tabu}
}
    \caption{Free parameters of the \shw\ model, their best-fitting values, and the associated uncertainties which are derived as per \cref{app:corner}. Note that \(\theta^\prime, \phi^\prime, \psi^\prime\) are derived from the best-fitting \(q,p,u\).}
    \label{tab:FP}
\end{table}
The \shw\ model provides a plethora of intrinsic properties, including the \ND{3} shapes of the mass distribution (defined by the parameters in \cref{tab:FP}) and velocity ellipsoid (defined by the specific orbit superposition). One of the important quantities for the test sought here is the projected mass enclosed within a circular aperture of radius \(\Rein\). We compute this in the following way. Each spatial aperture of the kinematic data has an associated luminosity weighting, which is the summation over the weights of all orbits which cross that aperture. We converted the luminosity weights into total mass weights by adding an MGE parametrisation of the best-fitting NFW halo to \mgeT, then dividing by \mgeS. This scale factor was evaluated at every aperture so that we could compute the fractional dynamical mass from the model. The projected fractional mass weight within \(\Rein\), \(f_{\rm Ein.}\), was then computed by simply summing all apertures with circular (projected) radius \(r \leq \Rein\). Then the enclosed dynamical mass was computed by combining and integrating the enclosed mass profiles of the stars, DM, and SMBH. We integrated these profiles up to the maximum radial extent of the kinematic data, \(r_{\rm max}\), producing an enclosed dynamical mass within the FOV, \(M_{\rm FOV}\). Finally, the projected dynamical mass within \(\Rein\) is given straight-forwardly as \(M_{\rm Ein.} = f_{\rm Ein.} M_{\rm FOV}\). The intrinsic enclosed mass profiles, along with the projected measurements from both the dynamical and lensing models, are presented in \cref{img:profile}.\par
From the inset of \cref{img:profile}, it is clear that the measurements of the projected Einstein mass from the lensing and dynamical models are in excellent agreement with one another, with \((9.49\pm 0.15)\times 10^{10}\ \si{\Msun}\) and \((\dynEinMass)\times 10^{10}\ \si{\Msun}\), respectively. The uncertainty on the result from the dynamical model is computed by repeating the measurement for every model evaluation within \(1\sigma\) of the best-fit solution, and finding the variance. While the dynamical mass is expected to be a robust quantity, this outcome is still reassuring in light of the generality and complexity of the \shw\ model, and given that the lensing and dynamical masses were completely independent of one another. By establishing that the \shw\ model is accurately anchored to the lensing result at \(\Rein\), we can then confidently explore the other properties and spatial regions of the model in more detail. The intrinsic enclosed mass profiles shown in \cref{img:profile} illustrate that SNL-1 is completely baryon-dominated within \(\Rein\). Enclosed within \(r_{\rm max}\), we measure a total stellar mass of \(\logM[\star] = \logsMassFOV\), and a dynamical mass of \(\logM[{\rm dyn.}] = \logtMassFOV\). Enclosed within \(R_\eff\), we measure \(\logM[\star] = \logsMassRe\) and \(\logM[{\rm dyn.}] = \logtMassRe\), implying a DM fraction \(f_{\rm DM}(r\leq R_\eff) = \fDMRe\%\). It is clear that SNL-1 is baryon-dominated within the spectroscopic FOV.\par
\begin{figure}
    \centerline{\includegraphics[width=\columnwidth]{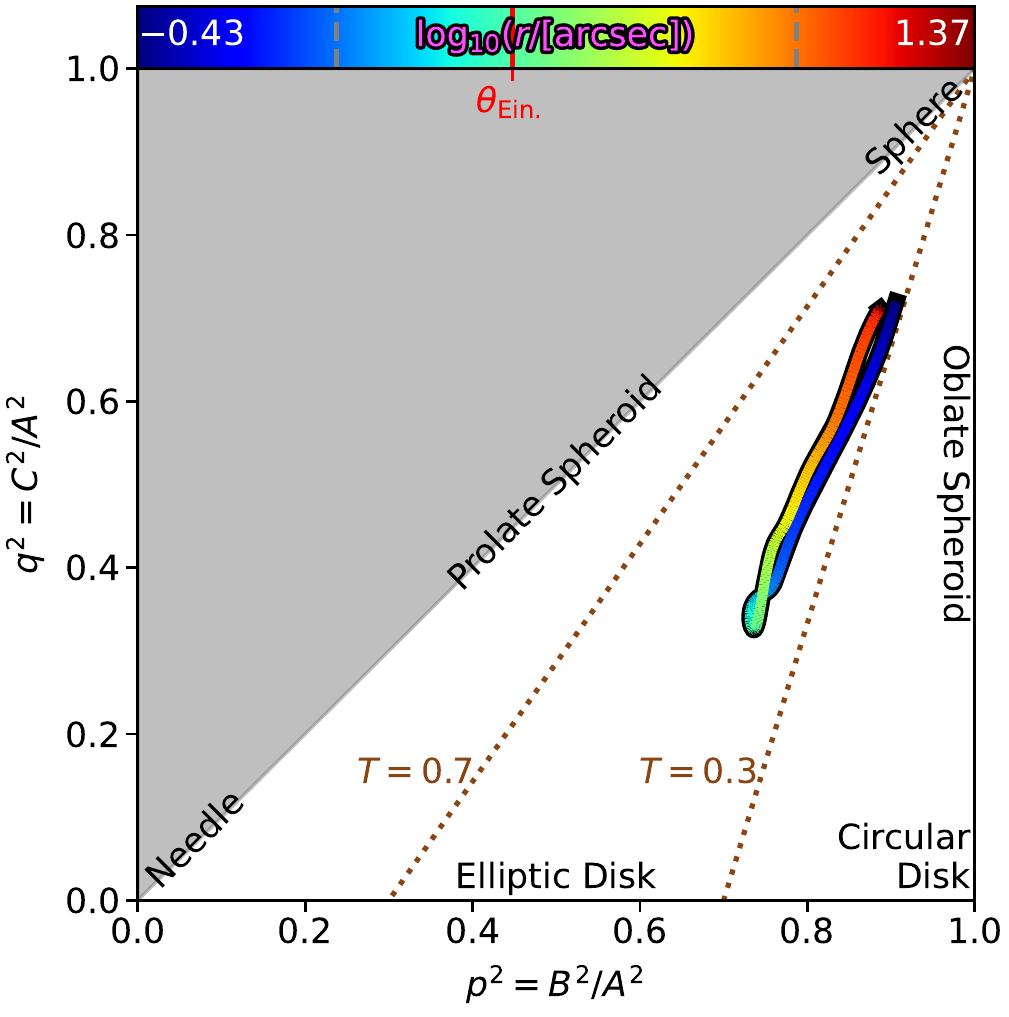}}
    \caption{Correlations of the axis ratios of the intrinsic stellar mass distribution. The curve is coloured by radius. The PSF and largest radial extent of the MUSE data are demarcated on the colour bar by the dashed grey lines. \(\Rein\) is also marked on the colour bar. The dotted brown lines illustrate the boundaries between the classes of triaxiality in \protect\cite{jin2020}. The various shape labels are borrowed from \protect\cite{dezeeuw1989a}. The radial profile illustrates that SNL-1 tends towards a spherical mass distribution at both small and large radii, but exhibits the strongest departure at \(\Rein\). Mild, approximately-constant triaxiality is present throughout the galaxy.}
    \label{img:axratio}
\end{figure}

The intrinsic shape of SNL-1 is explored in \cref{img:axratio} as a function of radius. This figure shows the radial dependence of the correlation between \(p^2\) and \(q^2\). In this way, the gradient of the curve is also approximately related to the triaxiality parameter \citep{franx1991}, \(T\equiv (1-p^2)/(1-q^2)\).
Strictly, our model of SNL-1 is triaxial at all radii. However \(p=B/A \gtrsim 0.9\) for all radii implying only minor triaxiality. \cite{jin2020} quantitatively categorises galaxies as oblate-triaxial, triaxial, and prolate-triaxial for \(T<0.3\), \(0.3 < T < 0.7\) and \(0.7 < T\), respectively. The mean triaxiality within the FOV for our model of SNL-1 is \(\left\langle T\right\rangle = \meanTriFOV\), implying that is it oblate-triaxial. Interestingly, \(q\) is smallest around \(\Rein\). This also happens to be the region in which the stellar rotation is high, and regularly-rotating galaxies are expected to be oblate \citep{weijmans2014}, so the value of \(q\) is perhaps unsurprising. The outskirts become more spheroidal, coincident with where the rotation diminishes. This could be the result of isotropic minor accretion. Galaxies as massive as SNL-1 are expected to have accreted a large portion of their present-day mass \citep{oser2010, khochfar2011, lackner2012, rodriguez-gomez2016}, which would preferentially settle in the outskirts \citep{karademir2019} and explain both the intrinsic shape and lack of rotation.\par
\begin{figure}
    \centerline{\includegraphics[width=\columnwidth]{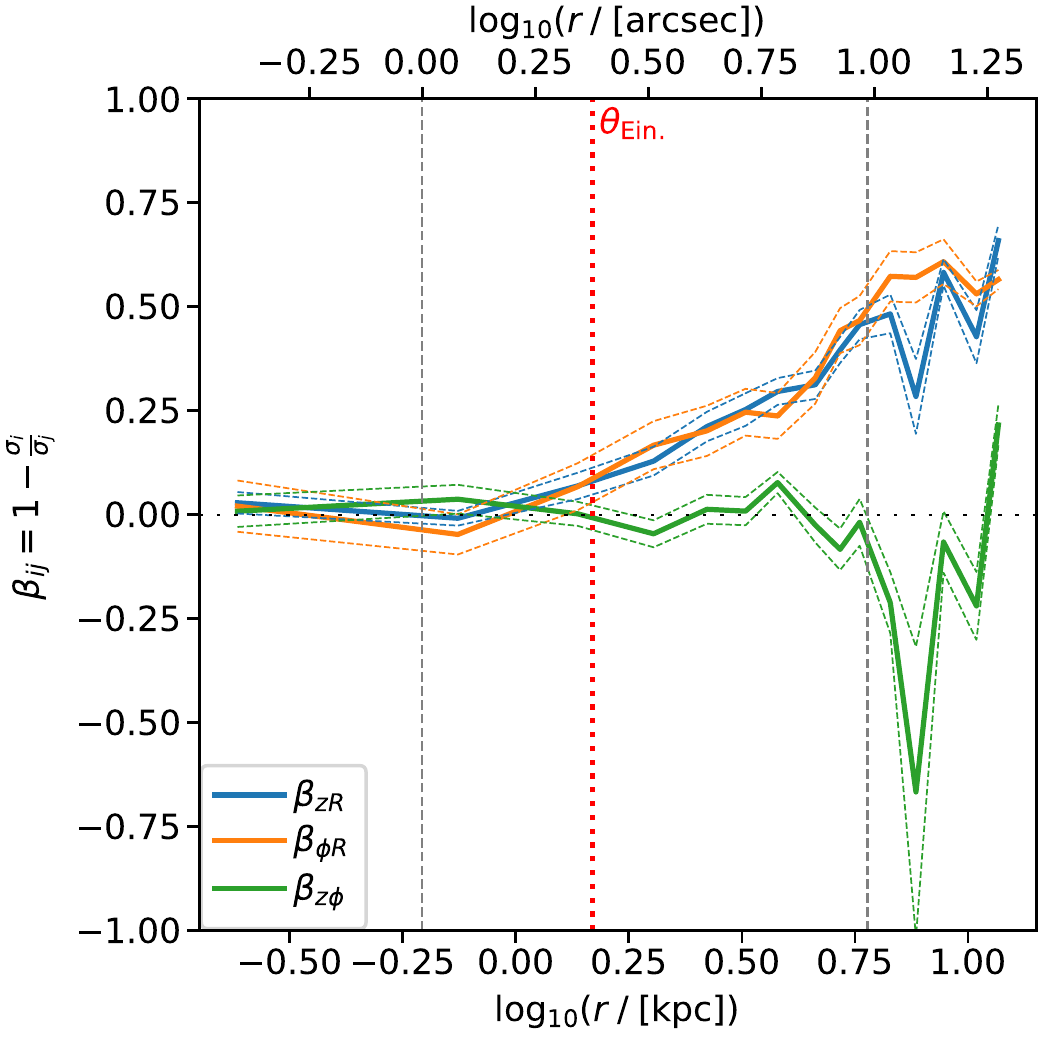}}
    \caption{Radial profiles of the intrinsic axes of the stellar velocity ellipsoid. The \(\beta_{ij}\) quantities are defined in \protect\cref{eq:anisotropy}. The PSF and largest radial extent of the MUSE data are demarcated by the inner and outer vertical dashed lines, respectively. \(\Rein\) is shown as the red dotted line. The coloured dashed lines illustrate the spread of all models within \(1\sigma\) of the best-fitting solution. SNL-1 is approximately isotopic in the central regions \((r < \Rein)\), while the outskirts are characterised by an intrinsically-oblate and anisotropic SVE.}
    \label{img:sve}
\end{figure}

Of further interest is the shape of the intrinsic stellar velocity ellipsoid (SVE). This is particularly important because some dynamical mass estimators --- which may also be used in joint lensing analyses --- are valid only under the assumption of orbital isotropy. Orbital anisotropy has, however, been indirectly investigated in some lensing studies \citep[e.g.][]{koopmans2009}. To test this assumption, we investigate the radial variations of anisotropy, for each pair of axes of a cylindrically-aligned SVE. We explore the classic anisotropy parameter \citep{binney1987}
\begin{equation}\label{eq:anisotropy}
    \beta_{ij} = 1 - \frac{\sigma_i}{\sigma_j}
\end{equation}
for every \((i,j) \in [R, \phi, z]\). These results are shown in \cref{img:sve}.
Within \(\Rein\), SNL-1 may be considered approximately isotropic; \(\sigma_R \approx \sigma_\phi \approx \sigma_z\) for \(r < \Rein\). At larger radii, SNL-1 becomes increasingly radially biased with increasing radius; \(\sigma_z\approx \sigma_\phi < \sigma_R\) for \(r>\Rein\). Such anisotropy provides additional evidence of a rich accretion history over which many systems contributed to the build-up of the outer spheroid \citep[e.g.][]{naab2006}, as has also been found in the Milky-Way \citep[e.g.][]{helmi2018}. It also indicates that within \(\Rein\), the assumption of orbital isotropy --- at least for SNL-1 --- would be reasonable.

\section{Discussion}
The dynamical model has afforded us a detailed look at the resolved internal kinematic properties of SNL-1. We see that the central regions exhibit high rotation, an oblate mass distribution, and approximately-isotropic orbits. The outskirts exhibit little-to-no rotation, a spheroidal mass distribution, and are accompanied by an increase in radially-biased orbits. Intermediate triaxiality persists throughout the galaxy.
\subsection{SNL-1 in Context}
Apart from being a strong lens, SNL-1 exists within the population of massive elliptical galaxies. It is, however, highly compact for its mass \citep{campbell2014, smith2015a, spiniello2015c}, and therefore may not be representative of the ETG population as a whole. We thus seek to compare it to similar galaxies in order to discern any differences between them. Exploiting the generality of the \shw\ model, we can make this comparison for a number of key intrinsic properties.\par
\begin{figure}
    \centerline{\includegraphics[width=\columnwidth]{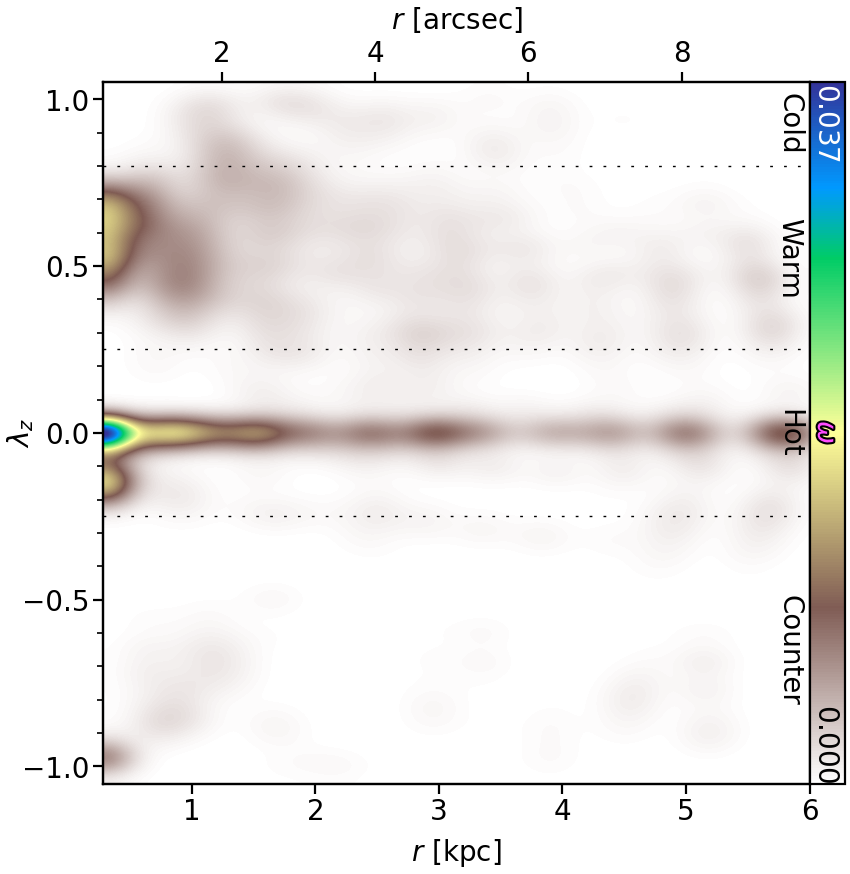}}
    \caption{Circularity distribution of the best-fitting \shw\ model within the spectroscopic FOV. The colour shows the orbital weight as a function of circularity and radius. The black dashes lines show the separation into four broad dynamical components (see text). SNL-1 contains predominantly dynamically-hot orbits at all radii. The inner rotating `disk' is visible for \(r \lesssim 2\si{\kilo\parsec}\) and \(\lambda_z \sim 0.6\). There is also the suggestion of counter-rotating orbits, though these contribute little mass.}
    \label{img:circ}
\end{figure}
\begin{figure}
    \centerline{\includegraphics[width=\columnwidth]{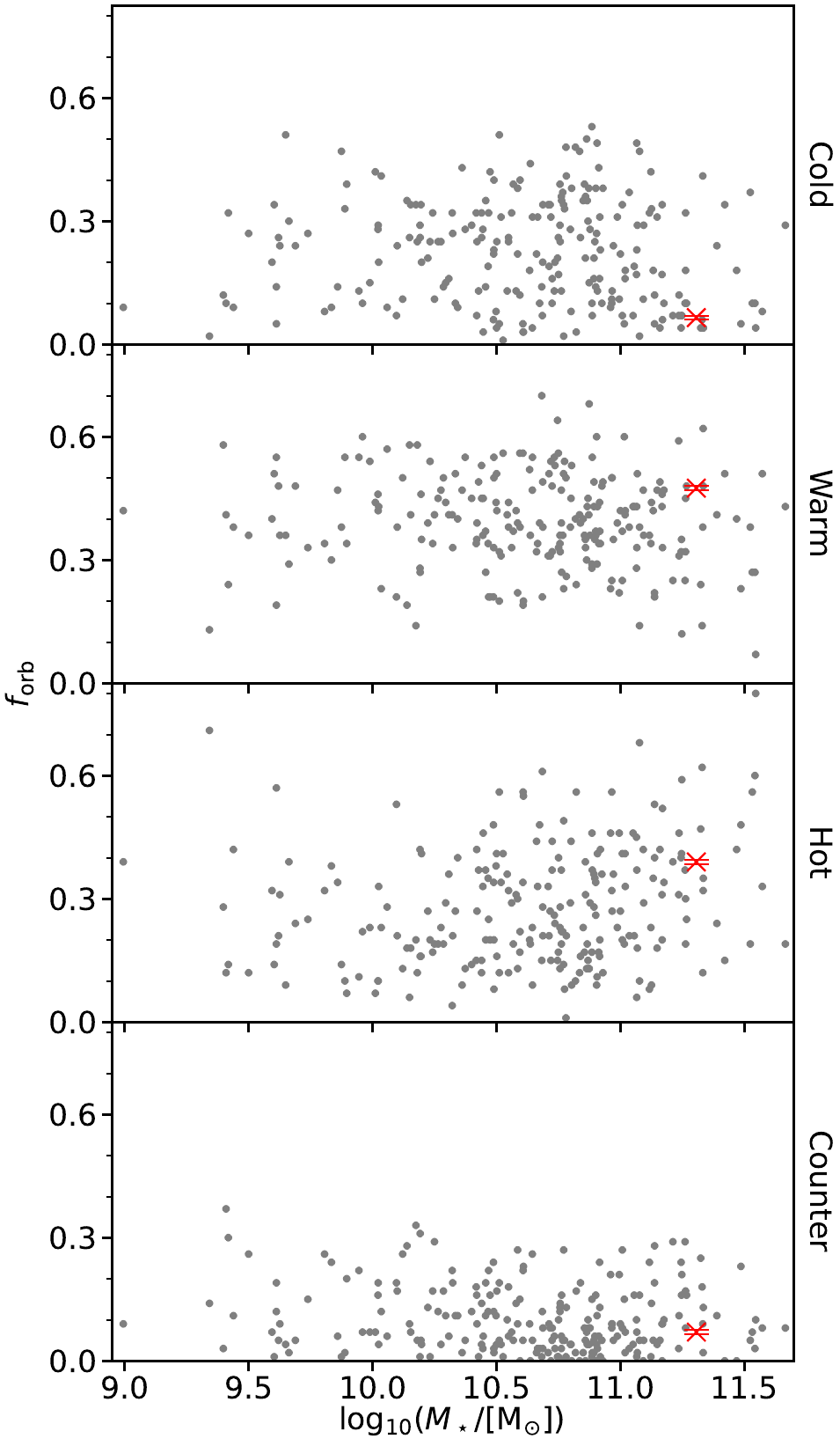}}
    \caption{Orbit fractions as a function of stellar mass. The grey dots show the fractions of cold, warm, hot, and counter-rotating orbits (top to bottom) from \protect\cite{zhu2018c} for the CALIFA sample. The corresponding values from the dynamical model of SNL-1 are marked by the red crosses. The illustrated uncertainty is the spread of all models within \(1\sigma\) of the best-fit.}
    \label{img:orbFrac}
\end{figure}

In \cref{img:circ}, the orbital circularity distribution is shown.
The orbital circularity \(\lambda_z\) is a measure of the intrinsic angular momentum around the short axis of the mass distribution \citep{zhu2018a}. It is normalised by the angular momentum of a circular orbit with equivalent energy, such that \(\lambda_z \in [-1, 1]\), representing counter- and co-rotating circular orbits, respectively. Box and/or radial orbits will exhibit \(\lambda_z \sim 0\). \cref{img:circ} shows that SNL-1 is composed of dynamically-hot and -warm orbits with \(\lambda_z \lesssim 0.7\). We use this distribution to compare to the analysis conducted on the representative sample of nearby galaxies from the CALIFA survey \citep{sanchez2012}, which contains a small number of galaxies of similar mass to SNL-1. \cite{zhu2018c} applied triaxial \shw\ models to that sample and analysed the distribution of orbits as a function of stellar mass by dividing the circularity into broad bins of `cold' \((\lambda_z \geq 0.8)\), `warm' \((0.25 < \lambda_z < 0.8)\), `hot' \((\left|\lambda_z\right| \leq 0.25)\), and counter-rotating \((\lambda_z < -0.25)\) orbits. They specifically considered the fractions of each orbit type within \(R_\eff\). Applying these same criteria, we compare orbits fractions to the CALIFA galaxy sample in \cref{img:orbFrac}.
The model of SNL-1 reveals cold, warm, hot, and counter-rotating fractions within \(R_\eff\) of \(\fOrbCold, \fOrbWarm, \fOrbHot,\text{ and }\fOrbCR\), respectively. SNL-1 is consistent with the \(\logM[\star] \sim 11\) population of galaxies from CALIFA, in terms of the orbit distributions. Given its mass, SNL-1 is expected to be supported predominantly by random motions \citep[for instance, see][]{cappellari2013}, and we see this borne out of our model as a significant fraction of hot orbits at all radii. Simultaneously, SNL-1 exhibits relatively high peak rotation velocity. It can be seen from \cref{img:circ} that this rotation is produced predominantly by the cloud of dynamically-warm orbits extending from the centre to \(\sim 2 \si{\kilo\parsec}\) (since there is negligible contributions from cold orbits), thereby explaining the relatively high warm-orbit fraction within \(R_\eff\). These orbits would constitute a `thick-disk'-like component, embedded in the otherwise spheroidal galaxy.\par
\begin{figure}
    \centerline{\includegraphics[width=\columnwidth]{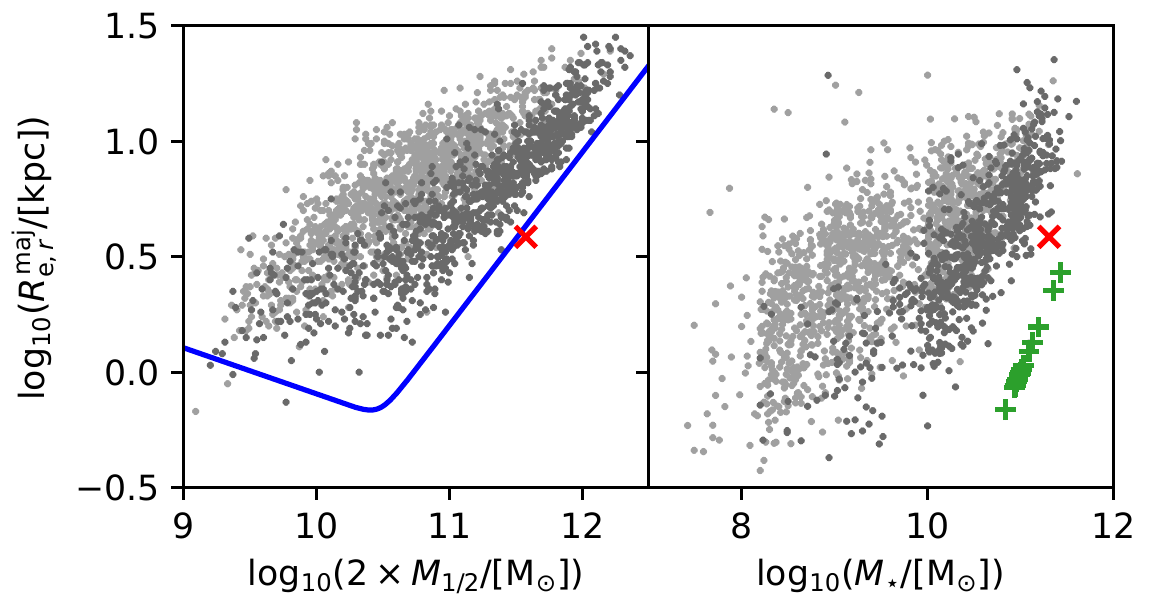}}
    \caption{Mass--size plane projections for two different literature samples. {\em Left:} the dynamical mass and the effective radius along the major axis \(R_\eff^{\rm maj}\) for the {\rm MaNGA} sample \protect\citep{li2018}. \(M_{1/2}\) is the dynamical mass enclosed within the \ND{3} spherical half-light radius \(R_{1/2}\). The blue line shows the ``zone of exclusion'' of \protect\cite{cappellari2013a}. {\em Right:} the total stellar mass and the effective radius along the major axis \(R_\eff^{\rm maj}\) for the {\rm GAMA} subset (non-cluster galaxies) of the SAMI sample \protect\citep{croom2021}. The green plus symbols show the sample of `relic' galaxies from the INSPIRE survey \protect\citep{spiniello2021a}. In both panels, the samples are coloured by morphological type; late-type galaxies are light grey (`S' from {\rm MaNGA} and \tfo{VisualMorphologyDR3.TYPE} \(\geq 2.0\) for SAMI), and earlier morphological types are dark grey. In both panels, the corresponding measurement from the dynamical model of SNL-1 is marked by the red cross. The sizes from both surveys are \(r\)-band measurements. We compute an \(r\)-band effective radius for SNL-1 from {\rm FORS2} photometry of \(R_{\eff, r}^{\rm maj} = 5.98\ \si{\arcsecond}\ (3.83\ \si{\kilo\parsec})\). In both panels, SNL-1 lies at the compact, high-mass edge of the general galaxy populations.}
    \label{img:massSize}
\end{figure}

In contrast to its rather typical dynamical properties, SNL-1 appears to be somewhat unusual structurally, being highly compact for its mass. We therefore investigate its position on the mass--size plane, an empirical correlation believed to capture the various evolutionary stages of a population of galaxies \citep[e.g.][]{newton2011, cappellari2013, cappellari2013b, scott2017, krajnovic2018b, li2018}.
%\citep[with any redshift evolution of this plane being negligible between SNL-1 and \(z=0\);][]{mcintosh2005, dutton2010, vandesande2011, vanderwel2014}.
In \cref{img:massSize}, we place SNL-1 on both the dynamical-mass--size plane measured in \cite{li2018} using the {\rm MaNGA} survey \citep{bundy2015}, and the stellar-mass--size plane using DR3 \citep{croom2021} of the SAMI survey \citep{croom2012}, comparing also to a sample of `relic' galaxies from the INSPIRE survey \citep{spiniello2021a}.
SNL-1 occupies the boundary at the compact, high-mass edge of the `normal' galaxy populations. It resides in the ``zone of exclusion'', which was derived dynamically for the \atd\ sample of galaxies in the Virgo cluster \citep[][but proposed originally by \protect\citealt{bender1992, burstein1997}]{cappellari2013b}. This may be the result of the fact that more compact galaxies are preferentially easier to detect as being strong lenses, as already hinted to in the results of \cite{newton2011}. Moreover, SNL-1 appears to bridge the gap between the relic and non-relic populations, the former being defined specifically by their compactness. This further reinforces the atypical compactness of SNL-1 with respect to the broader galaxy population.\par
Overall this section has shown that SNL-1 is dynamically typical but structurally atypical compared to galaxies of similar mass. Of particular interest for lensing models which utilise the central velocity dispersions as dynamical tracers, at its Einstein radius SNL-1 exhibits the greatest departure from a spherical mass distribution, but no significant departure from orbital isotropy. These results suggest that orientation/configuration biases are likely already manifest in existing lensing samples, while our methodology --- specifically the sophisticated orbit-based dynamical models --- provides a way to quantify and account for such biases in future samples. In this case, the properties derived from lensing analyses, even if large samples are collected, may not be representative of the broader galaxy population.

\subsection{Dynamical Analyses of Lensing Systems}\label{ssec:lensSys}
The lensing selection function is of considerable importance, and depends on the projected mass of the lens along the LOS \citep[for which the stellar velocity dispersion is a common proxy;][]{treu2010}. However, there are other physical properties of the lensing galaxies which would be especially conducive to producing detectable strong-lens signals. For instance, \cite{schaller2015a} studied a sample of galaxy clusters in cosmological hydrodynamical simulations and find that, of their \(6\) clusters, \(4\) brightest cluster galaxies are prolate, and \(2\) are oblate. They argue that if such galaxies were intrinsically prolate, but were selected for observation in such a way which favours end-on orientations, then the measured central velocity dispersion would be biased to higher values compared to a sample with a uniform distribution of orientations. A similar argument could apply in galaxy-scale lens systems, where internal structures such as bars could also play a role. Therefore, using the central velocity dispersion measurements as part of lens models would propagate this systematic bias and render the lensing sample non-representative. It is thus important to determine whether the selection function for lensing does indeed favour certain geometric configurations, and whether these configurations can be accurately reproduced in lens models.\par
%Since our models allow us to explicitly measure the viewing angles, we can begin to ascertain if such a bias is exacerbated in strong-lensing samples --- that is, if there are preferred viewing angles along which lensing signals are easier to detect.\par
It is to this end that dynamical models, such as those presented in this work, can provide insight. For example, the \(u\) parameter in this implementation of the \shw\ model describes the projection of a triaxial system. If strong lens galaxies have a tendency to be end-on triaxial systems, the distribution of \(u\) for a sample of lens galaxies will have a lower mean with respect to a mass-matched sample of non-lensed galaxies. \shw\ implementations which include treatment of figure rotation such as {\sc Forstand} \citep{vasiliev2021} could simultaneously constrain the orientation of time-variable triaxial structures such as bars. However, as prefaced in \cref{sec:intro}, the sample of strong lens galaxies with spatially-resolved stellar kinematics of sufficient quality is currently too small for a statistical analysis. Exploiting the adaptive-optics capabilities of upcoming facilities such as {\rm MAVIS} \citep{mcdermid2020} and {\rm HARMONI} \citep{thatte2016} will bring more distant strong lens galaxies into the regime amenable to this analysis.\par
Although a direct statistical test is not yet available, the existing mass models used by lensing analyses appear to agree with independent techniques. For instance, results from lensing show that the total-mass distributions of the lensing population exhibit a surprisingly strong tendency towards isothermal radial profiles --- having \(\gamma\sim -2.1\) for \(\rho_{\rm tot}(r)\propto r^\gamma\) \citep[e.g.][]{treu2004, koopmans2009, auger2010, barnabe2011, bolton2012, sonnenfeld2013, shajib2021}, with an intrinsic scatter of order \(0.1-0.2\) between them. This is in good agreement with direct modelling of the total mass distribution using Jeans models of the stellar kinematics \citep[][despite different internal assumptions between them]{tortora2014, cappellari2015, poci2017, bellstedt2018, li2019, derkenne2021}, cosmological simulations \citep{remus2013, wang2019}, and gas dynamical models \citep{serra2016}, at least for \(z\lesssim 0.5\) \citep{derkenne2021}. Since these studies are on physically similar but non-lensed galaxies, this could imply at face value that there is no additional lensing selection bias beyond the velocity dispersion. However, by construction, assuming a single spherical power-law for the total mass of galaxies \citep[or even an oblate total-mass distribution;][]{poci2017} can not account for triaxiality (or bars), and so this comparison can not exclude the existence of such a bias.\par
\begin{figure}
    \centerline{\includegraphics[width=\columnwidth]{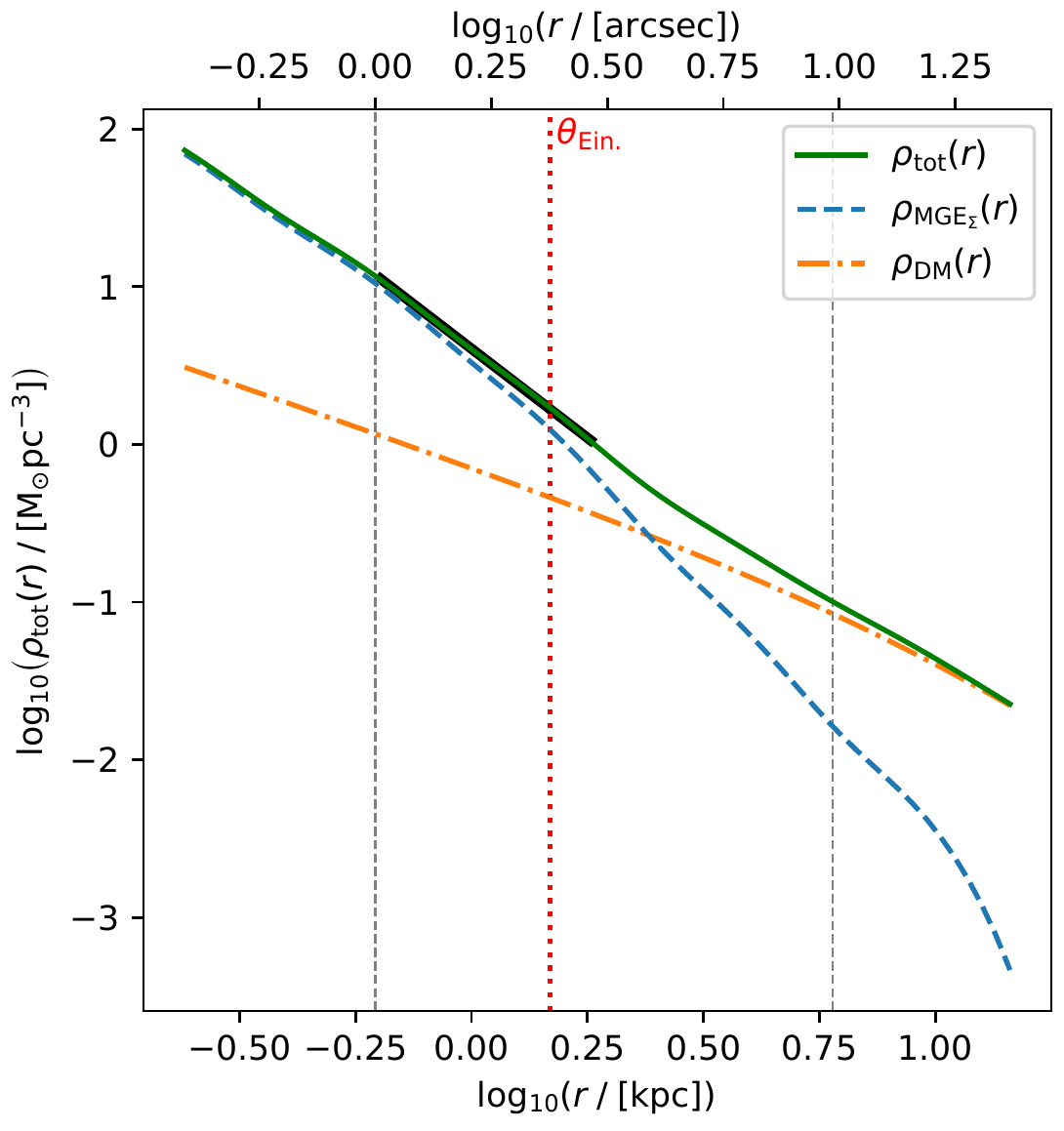}}
    \caption{Density profiles for SNL-1 from the best-fit \shw\ model. The total enclosed mass is shown in green [given by \(M_\star(r) + M_{\rm DM}(r)\)], the stellar mass is dashed blue, and the DM is dot-dashed orange. Underplotted in the thick black line is the single power-law \(\rho_{\rm tot}(r) \propto r^{\gamma^\prime}\) with the value of \(\gamma^\prime\) measured from \protect\cref{eq:pl}. SNL-1 does indeed have a mean logarithmic density slope which is close to isothermal over the fitted radial range. In addition, the black line illustrates that a power-law is a good representation of the total-mass density for SNL-1 over the region it is measured.}
    \label{img:massDens}
\end{figure}

We can calculate the total-mass profile slope from the dynamical model in this work by simply combining the best-fitting stellar and DM contributions. This is given in \cref{img:massDens}, presented on the same scheme as \cref{img:profile}. Given the flexibility of the models used in this work, there is no guarantee that the resulting total-mass density should be accurately described by a single power-law in radius. We therefore measure the mean logarithmic density slope within \(R_\eff\) given as
\begin{equation}\label{eq:pl}
\gamma^\prime = \frac{\log_{10}[\rho_{\rm tot}(R_\eff)/\rho_{\rm tot}(R_{\rm in})]}{\log_{10}(R_\eff/R_{\rm in})}
\end{equation}
where we set \(R_{\rm in} = 1\si{\arcsecond}\), the FWHM of the kinematic data.
We measure a mean logarithmic slope of the total-mass density profile to be \(\gamma^\prime = \totalSlope\) (\(1\sigma\) uncertainty), which is within the distributions found in the aforementioned works from lensing, stellar and gas kinematics, and cosmological simulations. While in this work we do not constrain the total-mass directly like previous studies, \cite{poci2017} showed that modelling the total-mass directly or considering stellar+DM contributions produce consistent results, while \cite{barnabe2011} showed that the slopes of the total-mass profiles between Jeans models and (two-integral) \shw\ models agree.\par
Finally more broadly, we have derived a measurement of \(M_{\rm Ein.}\) from the \shw\ model which is in excellent agreement with that from the independent lensing constraints. This result justifies being able to leverage lensing constraints directly within dynamical models. Doing so would be especially useful when constraints from the stellar kinematics alone are limited --- for instance, at higher redshift. Requiring any model to respect both the enclosed lensing mass and the kinematic constraints simultaneously would also allow degeneracies to be mitigated concerning implied gradients in \(M/L\), among others.

\subsection{Exploring Different Orientations}
\begin{figure}
\centerline{
    \includegraphics[width=\columnwidth]{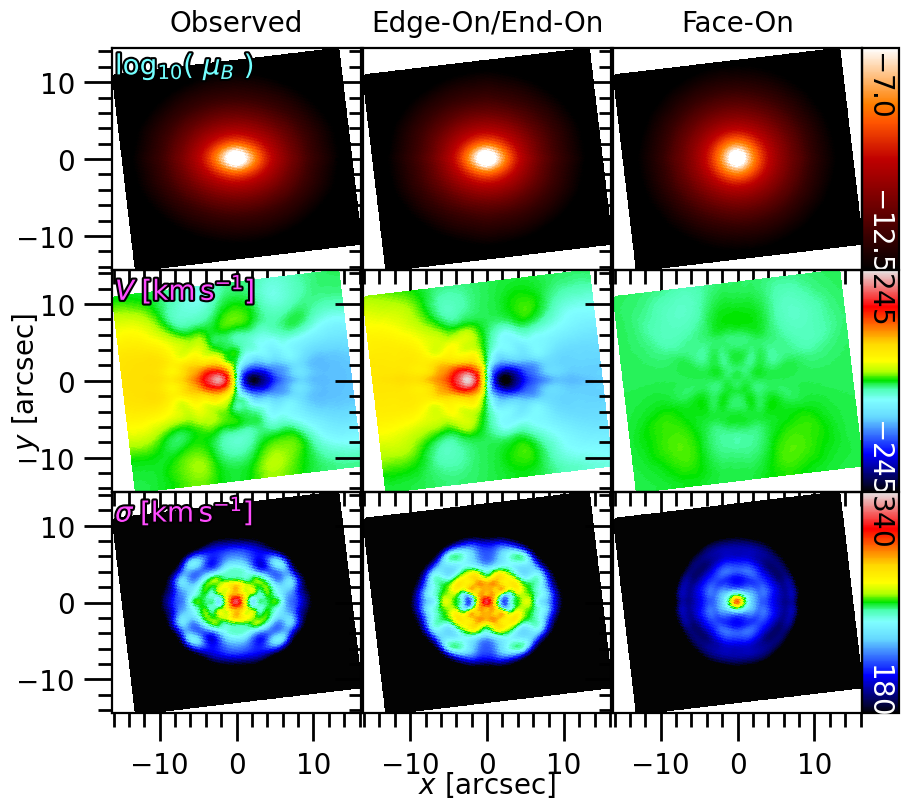}
}
\caption{Re-projected \shw\ models of SNL-1. From top to bottom are the projected surface brightness, mean velocity, and velocity dispersion. From left to right, the viewing angles are: observed \((\theta, \phi, \psi) = (\theta^\prime, \phi^\prime, \psi^\prime)\);  edge-on and end-on \((\theta, \phi, \psi) = (90\si{\degree}, 0\si{\degree}, \psi^\prime)\); face-on \((\theta, \phi, \psi) = (0\si{\degree}, 90\si{\degree}, \psi^\prime)\). The kinematics are shown simply for illustration, and align with the expectation for these viewing directions. The changes to the surface brightness (and correspondingly the mass density) allow us to probe the impact of particular viewing directions on the strength of the lensing signal.}
\label{img:reproj}
\end{figure}

The \shw\ model is fully defined in \ND{3} space, and is projected through the LOS only to compare to observations. We are therefore able to `observe' this model at arbitrary orientations. In this section we explore particular re-projections of the best-fit dynamical model, and their impact on the lensing properties. We show examples in \cref{img:reproj}.
These figures are constructed by taking the distribution of orbits and their relative weights from the original \shw\ model, then `observing' the model from the specified viewing direction. New LOSVD are computed from this viewing direction. The surface brightness and kinematics are then measured by integrating, and fitting a Gauss-Hermite function to, these LOSVD, respectively.
%Surface brightness is converted to mass density assuming a spatially-constant scale factor (see below).
Since the models are spatially binned for the sole purpose of comparing to observations, we explore these re-projections on a non-binned pixel grid, but maintaining the MUSE pixel scale of \(0.2\ \si{arcsec \per pixel}\).\par
We explore two projections in addition to the observed direction in \cref{img:reproj}, defined by their viewing angles: from left to right, \((\theta, \phi, \psi) = (\theta^\prime, \phi^\prime, \psi^\prime); (\theta, \phi, \psi) = (90\si{\degree}, 0\si{\degree}, \psi^\prime); (\theta, \phi, \psi) = (0\si{\degree}, 90, \psi^\prime)\). Primed angles refer to their observed values; that is, those derived from the best-fitting \shw\ model and given in \cref{tab:FP}. The additional projections correspond to what are expected to be the most and least dense along the LOS, and should therefore have the largest impact on the lensing cross-section. The kinematics exhibit changes which are expected for the different orientations; the edge-on projection shows the highest amplitude of rotation, which effectively vanishes in the face-on projection. The velocity dispersion also contains the signature of high rotation in the edge-on projection, with a depression along the mid-plane.\par
\begin{figure*}
    \centerline{\includegraphics[width=\textwidth]{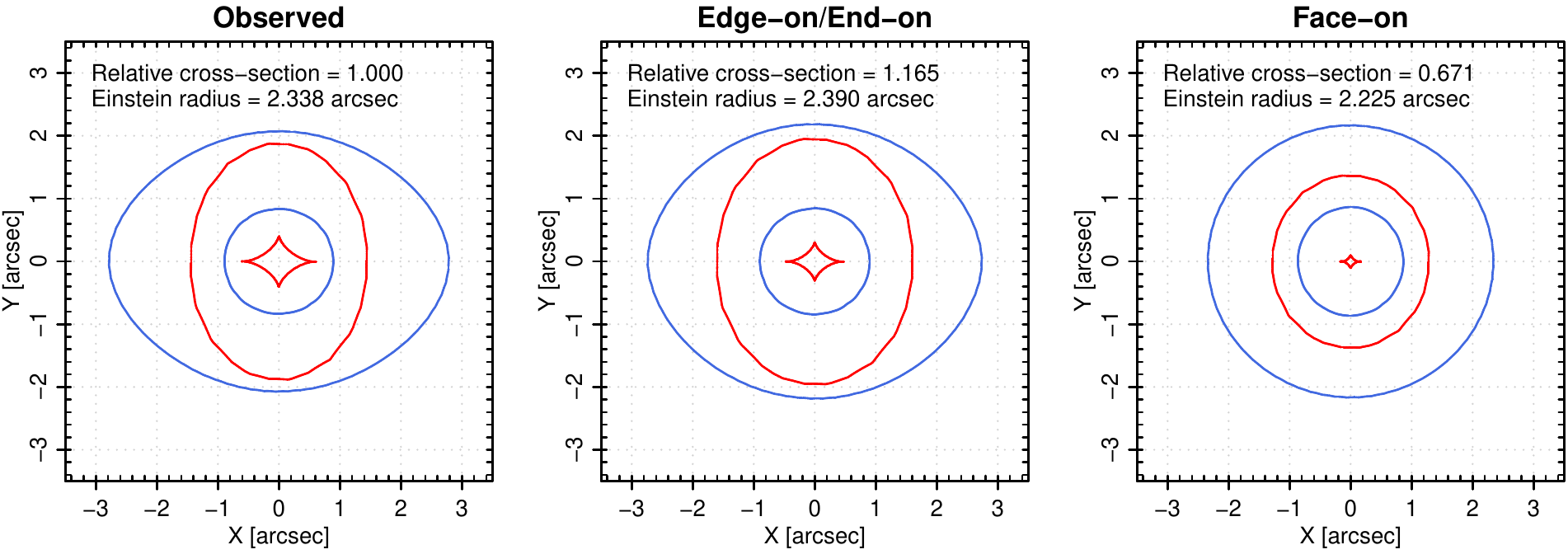}}
    \caption{Lensing properties of SNL-1 for three different orientations, as computed from the re-projected dynamical models. From left to right is the observed, edge-on/end-on, and face-on projections. The surface densities have been rescaled to lensing convergence by a single constant factor which reproduces the observed Einstein radius in the observed orientation. The blue and red lines show the image-plane critical curves and source-plane caustics, respectively. The cross-section for lensing is given by the area inside the outer caustic: the cross section in the observed orientation is close to that of the ``maximal-lensing'' edge-on/end-on orientation, and substantially larger than the ``minimal-lensing'' face-on case.}
    \label{img:lensCross}
\end{figure*}

%The elongation (seen in the face-on projection of the velocity dispersion) may be the result of the apparent lop-sidedness seen most clearly in the observed velocity dispersion map in \cref{img:schw}. In order to reproduce this feature in projection, the model would need to introduce an elongated structure at an angle to the observed LOS. We tested the significance of the elongation by probing the good-fitting models in the parameter-space of \cref{img:corner} neighbouring the best-fit solution, and it is present in all such models. The observed velocity dispersion map is, however, noisy in the region where the lop-sidedness occurs. We therefore conservatively avoid speculating further on the physical significance of this feature.\par
From the surface brightness of each projection, we compute the lensing cross-section by assuming a spatially-constant \mlStar, and neglecting the small DM contribution within \(\Rein\) (discussed below). These are shown in \cref{img:lensCross}.
The face-on and edge-on/end-on projections exhibit a cross-section area which is \(\faceOnCS\times\) and \(\edgeOnCS\times\) that of the observed projection. Conversely, we see only a \(\sim 10\%\) change in \(\Rein\) between the extrema. The edge-on/end-on relative cross-section is perhaps unsurprising, since the best-fit dynamical model has a corresponding inclination of \(\sim 88\si{\degree}\). This implies that SNL-1 is already in a configuration which is close to its `maximum' projection, and suggests that it was detected as a lens in SNELLS at least partially due to its favourable orientation. Moreover, the substantial change in lensing cross-section and central velocity dispersion for the different projections, with little change to \(\Rein\), might also produce different inferred mass profiles from lensing techniques which utilise the central kinematics \citep[e.g.][]{treu2010}.\par
Applying a spatially-constant \mlStar\ may impact the projected mass measurements for each orientation. We of course do not have access to the \mlStar\ integrated along lines-of-sight other than the observed orientation. However, the radial geometry of the observed \mlStar\ map (\cref{img:ML}, inset) indicates that this geometry may hold for all projections. In that case, and since we are measuring the cross-sections relative to the observed orientation, applying the same \mlStar \((R)\) profile (such as the one measured in \cref{img:ML}) to each projection will produce the same relative differences as a simple spatially-constant \mlStar. Similarly, given the spherical NFW employed in the dynamical model, the DM also does not have an impact on the relative measurements for the re-projections.

\section{Conclusion}
We have explored general orbit-based dynamical models of the relatively-nearby strong-lens system SNL-1. We have compared our results directly to the measurements from lensing models, and investigated a host intrinsic properties. Our findings are summarised here.
\begin{itemize}
    \item The enclosed mass measurements between the dynamical and lensing models agree well within their respective uncertainties, with \(M_{\rm Ein.}^{\rm lens} / M_{\rm Ein.}^{\rm Schw.} = \lensDynMassRatios\). This result supports the conclusion from the lensing model which favoured a relatively dwarf-poor IMF such as \cite{kroupa2001} for the centre of SNL-1 \citep{smith2015a}. We found that SNL-1 is baryon-dominated within the spectroscopic FOV, which encompasses both \(\Rein\) and \(R_\eff\) (\cref{img:profile}).
    \item SNL-1 appears to be oblate-triaxial and radially-anisotropic in its outer regions (\cref{img:axratio,img:sve}). In general, models (lensing or otherwise) which attempt to separate the baryonic and DM contributions to the total mass need to be able to account for these physical properties.
    \item Dynamically, SNL-1 is consistent with the broader galaxy population at fixed mass as traced by the CALIFA survey. Despite appearing to be a typical old, massive, red ETG, SNL-1 still exhibits relatively high rotation velocities in a central thick disk-like configuration. This is evidenced by the somewhat high proportion of kinematically-warm orbits within \(R_\eff\) (\cref{img:orbFrac}).
    \item In contrast, SNL-1 is markedly more compact at fixed mass compared to the {\rm MaNGA} and (field) {\rm SAMI} samples (\cref{img:massSize}). This compactness likely enhances the lensing signal compared to that of lower-density galaxies even at fixed mass. Nevertheless, the shape of the mass density profile of SNL-1 is in good agreement with a wide range of galaxy samples (\cref{img:massDens}).
    \item Finally, we explored the impact of changing the observed orientation of a galaxy which is already known to be a lens. We find a substantial impact on the lensing cross-section of a factor of \(\sim 2\) between the projected extrema. We find only a minor difference between the observed and maximal orientations, implying that the selection of SNL-1 as a lens may have been impacted by its observed orientation (\cref{img:lensCross}).
\end{itemize}
We conclude that combing lensing and sophisticated orbit-based dynamical models will provide access to the intrinsic physical properties of galaxies in a robust manner. In anticipation of the increase in the number of suitable galaxies expected with upcoming facilities, we continue to investigate the ways in which lensing and dynamical analyses can complement one another.

\section*{Acknowledgements}
Both authors are supported by the Science and Technology Facilities Council through the Durham Astronomy Consolidated Grant 2020–2023 (ST/T000244/1). This research is based on observations collected at the European Organisation for Astronomical Research in the Southern Hemisphere under ESO programme 0100.B-0769(A). This work used the DiRAC\@Durham facility managed by the Institute for Computational Cosmology on behalf of the STFC DiRAC HPC Facility (www.dirac.ac.uk). The equipment was funded by BEIS capital funding via STFC capital grants ST/K00042X/1, ST/P002293/1, ST/R002371/1 and ST/S002502/1, Durham University and STFC operations grant ST/R000832/1. DiRAC is part of the National e-Infrastructure. This work also made use of the \tfo{OzStar} supercomputer at Swinbourne University. This work utilised existing software packages for data analysis and presentation, including \tso{AstroPy} \citep{astropycollaboration2013}, \tso{Cython} \citep{behnel2011}, \tso{IPython} \citep{perez2007}, \tso{matplotlib} \citep{hunter2007}, \tso{NumPy} \citep{harris2020a}, the \tso{SciPy} ecosystem \citep{virtanen2020}, and \tso{statsmodels} \citep{seabold2010}. We finally thank the anonymous referee for comments which improved the clarity of this work.

\section*{Data Availability}
The observational data used in this work are publicly available in the
ESO archive (\url{archive.eso.org}). Other products can be provided
upon reasonable request to the author.

%%%%%%%%%%%%%%%%%%%%%%%%%%%%%%%%%%%%%%%%%%%%%%%%%%

%%%%%%%%%%%%%%%%%%%% REFERENCES %%%%%%%%%%%%%%%%%%

% The best way to enter references is to use BibTeX:

\bibliographystyle{mnras}
\bibliography{snl1Enclv6} % if your bibtex file is called example.bib

%%%%%%%%%%%%%%%%%%%%%%%%%%%%%%%%%%%%%%%%%%%%%%%%%%

%%%%%%%%%%%%%%%%% APPENDICES %%%%%%%%%%%%%%%%%%%%%
%
\appendix
\section{Projected Density Models}\label{app:MGE}
\cref{tab:mge} presents both the stellar mass density and luminosity density MGE models for SNL-1. For the models in this work, the MGEs are assumed to be axisymmetric in projection, and so all components have the same position angle.
\begin{table}
	\begin{tabular}{S[table-format=6.2]|S[table-format=6.2]|S[table-format=3.3]|c}
	 $\Sigma_\star$ & $\mu_B$ & $\sigma$ & $q$\\
	 $[\si{\Msun\per\parsec\squared}]$ & $[\si{\Lsun\per\parsec\squared}]$ & $[{\rm arcsec}]$ & \\
	\hline
665322.22 & 111570.70 & 0.022 & 0.700\\
241694.28 & 40530.74 & 0.075 & 0.500\\
66986.41 & 11235.65 & 0.093 & 0.900\\
109985.28 & 18455.82 & 0.212 & 0.900\\
21783.49 & 3657.34 & 0.491 & 0.900\\
43580.97 & 7328.28 & 0.612 & 0.503\\
21490.22 & 3619.68 & 1.279 & 0.500\\
6248.14 & 1053.47 & 1.396 & 0.812\\
3028.55 & 520.74 & 2.268 & 0.517\\
3128.43 & 551.26 & 3.426 & 0.740\\
712.15 & 135.44 & 8.168 & 0.846\\
	\hline
	\end{tabular}
    \caption{\mgeT\ and \mgeS\ for SNL-1. The columns represent, from left to right, the projected surface mass density, projected surface brightness, width (peak location), and axis ratio, respectively. By construction (\cref{ssec:mge}), the Gaussians for both MGE models have the same widths and axis ratios. Moreover, all Gaussian components have the same fixed PA.}
    \label{tab:mge}
\end{table}

\section{Schwarzschild Model Parameter-Space Exploration}\label{app:corner}
\begin{figure*}
    \centerline{\includegraphics[width=\textwidth]{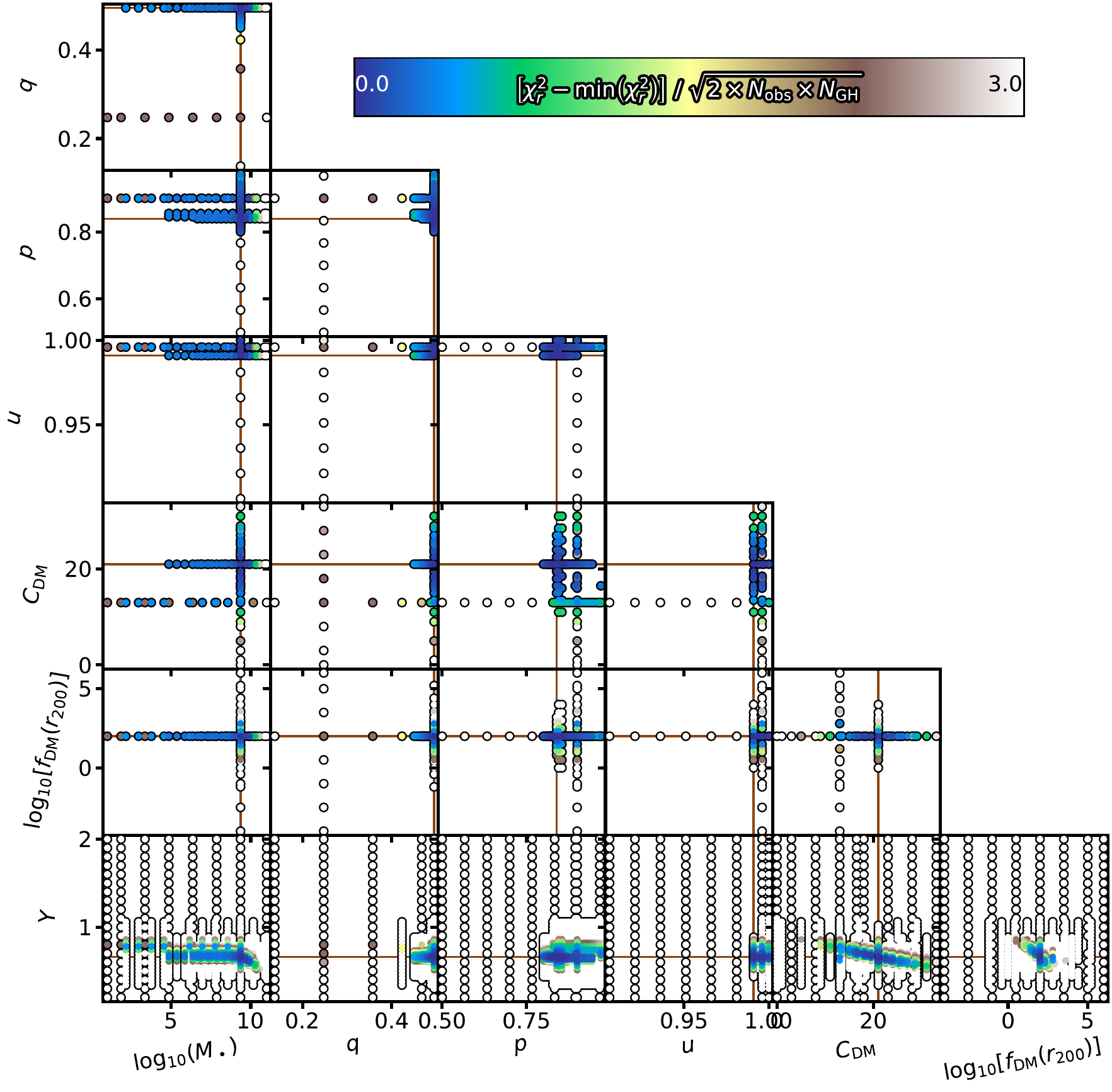}}
    \caption{\shw\ model parameter-space, showing the exploration of all free parameters. Each model is shown as a point, coloured by its reduced \(\chi^2\). The best-fit values are denote by the brown lines.}
    \label{img:corner}
\end{figure*}

Owing to the computationally-expensive implementations of the \shw\ technique, sampling methods such as Marko Chain Monte Carlo are currently impractical for model selection. The exploration of the parameter-space is instead achieved with a grid search over reasonable parameter ranges. In the case of the \(q\) shape parameter, its maximum value is constrained by the flattening of the MGE model \citep{vandenbosch2008}. The search is initialised over a wide range in each parameter to avoid local minima, then follows the minimum \(\chi^2\) while iteratively decreasing the step size. The search terminates once all surrounding models produce worse fits to the data. The search over the parameter-space is shown in \cref{img:corner}. This flexible grid approach still allows for the characterisation of the parameter-space, in particular any degeneracies between parameters, despite its relative simplicity.\par
The goodness-of-fit metric used here is a re-normalised version of the \(\chi^2\). The normalisation is a factor of \(\sqrt{2\times N_{\rm obs} \times N_{\rm GH}}\), where \(N_{\rm obs}\) is the number of spatial bins, and \(N_{\rm GH}\) is the number of kinematic moments fit by the model (four in this work; \cref{ssec:spec}) --- see \cite{zhu2018a} for further discussion on the goodness-of-fit. In order to estimate the uncertainties of the model parameters (\cref{tab:FP}), we take all models within \(1\sigma\) given by this normalised metric, and for each parameter compute the standard deviation marginalised over all other parameters.

\end{document}